\documentclass[aps,floatfix,prl,twocolumn,showpacs,superscriptaddress]{revtex4-1} 
\usepackage{longtable,dcolumn,epsfig,color}

\begin{document}
\def\nuc#1#2{${}^{#1}$#2}
\def\mee{$\langle m_{\beta\beta} \rangle$}
\def\mnu{$m_{\nu}$}
\def\ml{$m_{lightest}$}
\def\gnu{$\langle g_{\nu,\chi}\rangle$}
\def\mmod{$\| \langle m_{\beta\beta} \rangle \|$}
\def\mb{$\langle m_{\beta} \rangle$}
\def\BBz{$\beta\beta(0\nu)$}
\def\BBm{$\beta\beta(0\nu,\chi)$}
\def\BBt{$\beta\beta(2\nu)$}
\def\nonubb{$\beta\beta(0\nu)$}
\def\twonubb{$\beta\beta(2\nu)$}
\def\BB{$\beta\beta$}
\def\Mz{$M_{0\nu}$}
\def\Mt{$M_{2\nu}$}
\def\MzG{$M^{GT}_{0\nu}$} 
\def\MzF{$M^{F}_{0\nu}$} 
\def\MtG{$M^{GT}_{2\nu}$} 
\def\MtF{$M^{F}_{2\nu}$} 
\def\Gz{$G_{0\nu}$}					
\def\Tz{$T^{0\nu}_{1/2}$}
\def\Tt{$T^{2\nu}_{1/2}$}
\def\Tc{$T^{0\nu\,\chi}_{1/2}$}
\def\Rz{$\Gamma_{0\nu}$} 
\def\Rt{$\Gamma_{2\nu}$} 
\def\ms{$\Delta m_{\rm sol}^{2}$}
\def\ma{$\Delta m_{\rm atm}^{2}$}
\def\mot{$\Delta m_{12}^{2}$}
\def\mtt{$\Delta m_{23}^{2}$}
\def\ts{$\theta_{\rm sol}$}
\def\ta{$\theta_{\rm atm}$}
\def\ttwo{$\theta_{12}$}
\def\tot{$\theta_{13}$}
\def\gpp{$g_{pp}$} 
\def\gA{$g_{A}$} 
\def\qval{$Q_{\beta\beta}$} 
\def\be{\begin{equation}}
\def\ee{\end{equation}}
\def\cpKkgy{cnts/(keV kg y)}
\def\cpKkgd{cnts/(keV kg d)}
\def\cpRty{cnts/(ROI t y)}
\def\onecpRty{1~cnt/(ROI t y)}
\def\threecpRty{3~cnts/(ROI t y)}
\def\ppc{P-PC} 
\def\nsc{N-SC} 
\def\cosixty{$^{60}Co$}
\def\thttt{$^{232}\mathrm{Th}$}
\def\utte{$^{238}\mathrm{U}$}
\def\mubqkg{$\mu\mathrm{Bq/kg}$}
\def\cusulfate{$\mathrm{CuSO}_4$}
\def\MJ{{\sc Majorana}} 
\def\DEM{{\sc Demonstrator}} 
\def\MJDEMbf{\bfseries{\scshape{Majorana Demonstrator}}}
\def\MJbf{\bfseries{\scshape{Majorana}}}
\def\MJDEMit{\itshape{\scshape{Majorana Demonstrator}}}
\newcommand{\Gerda}{GERDA}
\newcommand{\GF}{\textsc{Geant4}}
\newcommand{\MaGe}{\textsc{MaGe}}

\title{First Limit on the Direct Detection of Lightly Ionizing Particles for Electric Charge as Low as $e$/1000 with the \MJ\ \DEM }

\newcommand{\ITEP}{National Research Center ``Kurchatov Institute'' Institute for Theoretical and Experimental Physics, Moscow, Russia}
\newcommand{\JINR}{Joint Institute for Nuclear Research, Dubna, Russia}
\newcommand{\lbnl}{Nuclear Science Division, Lawrence Berkeley National Laboratory, Berkeley, CA, USA}
\newcommand{\lanl}{Los Alamos National Laboratory, Los Alamos, NM, USA}
\newcommand{\queens}{Department of Physics, Engineering Physics and Astronomy, Queen's University, Kingston, ON, Canada} 
\newcommand{\uw}{Center for Experimental Nuclear Physics and Astrophysics,
and Department of Physics, University of Washington, Seattle, WA, USA}
\newcommand{\unc}{Department of Physics and Astronomy, University of North Carolina, Chapel Hill, NC, USA}
\newcommand{\duke}{Department of Physics, Duke University, Durham, NC, USA}
\newcommand{\ncsu}{Department of Physics, North Carolina State University, Raleigh, NC, USA}	
\newcommand{\ornl}{Oak Ridge National Laboratory, Oak Ridge, TN, USA}
\newcommand{\ou}{Research Center for Nuclear Physics, Osaka University, Ibaraki, Osaka, Japan}
\newcommand{\pnnl}{Pacific Northwest National Laboratory, Richland, WA, USA}
\newcommand{\princeton}{Department of Physics, Princeton University, Princeton, NJ, USA}
\newcommand{\sdsmt}{South Dakota School of Mines and Technology, Rapid City, SD, USA}
\newcommand{\usc}{Department of Physics and Astronomy, University of South Carolina, Columbia, SC, USA}
\newcommand{\usd}{Department of Physics, University of South Dakota, Vermillion, SD, USA} 
\newcommand{\ut}{Department of Physics and Astronomy, University of Tennessee, Knoxville, TN, USA}
\newcommand{\tunl}{Triangle Universities Nuclear Laboratory, Durham, NC, USA}
\newcommand{\mpi}{Max-Planck-Institut f\"{u}r Physik, M\"{u}nchen, Germany}
\newcommand{\tum}{Physik Department, Technische Universit\"{a}t, M\"{u}nchen, Germany}
\newcommand{\MIT}{Department of Physics, Massachusetts Institute of Technology, Cambridge, MA, USA} 

\affiliation{\uw}
\affiliation{\pnnl}
\affiliation{\usc}
\affiliation{\ornl}
\affiliation{\ITEP}
\affiliation{\usd}
\affiliation{\mpi}
\affiliation{\JINR}
\affiliation{\duke}
\affiliation{\tunl}
\affiliation{\uw}
\affiliation{\unc}
\affiliation{\lbnl}
\affiliation{\sdsmt}
\affiliation{\lanl}
\affiliation{\ut}
\affiliation{\ou}
\affiliation{\princeton}
\affiliation{\ncsu}
\affiliation{\MIT}
\affiliation{\queens}
\affiliation{\tum}

\author{S.I.~Alvis}\affiliation{\uw}
\author{I.J.~Arnquist}\affiliation{\pnnl} 
\author{F.T.~Avignone~III}\affiliation{\usc}\affiliation{\ornl}
\author{A.S.~Barabash}\affiliation{\ITEP}
\author{C.J.~Barton}\affiliation{\usd}
\author{F.E.~Bertrand}\affiliation{\ornl}
\author{V.~Brudanin}\affiliation{\JINR}
\author{M.~Busch}\affiliation{\duke}\affiliation{\tunl}	
\author{M.~Buuck}\affiliation{\uw} 
\author{T.S.~Caldwell}\affiliation{\unc}\affiliation{\tunl}	
\author{Y-D.~Chan}\affiliation{\lbnl}
\author{C.D.~Christofferson}\affiliation{\sdsmt} 
\author{P.-H.~Chu}\affiliation{\lanl}
\author{C. Cuesta}\altaffiliation{Present address: Centro de Investigaciones Energ\'{e}ticas, Medioambientales y Tecnol\'{o}gicas, CIEMAT 28040, Madrid, Spain}\affiliation{\uw}	\author{J.A.~Detwiler}\affiliation{\uw}
\author{C. Dunagan}\affiliation{\sdsmt}
\author{Yu.~Efremenko}\affiliation{\ut}\affiliation{\ornl}
\author{H.~Ejiri}\affiliation{\ou}
\author{S.R.~Elliott}\affiliation{\lanl}
\author{T.~Gilliss}\affiliation{\unc}\affiliation{\tunl} 
\author{G.K.~Giovanetti}\affiliation{\princeton} 
\author{M.P.~Green}\affiliation{\ncsu}\affiliation{\tunl}\affiliation{\ornl} 
\author{J. Gruszko}\affiliation{\MIT}
\author{I.S.~Guinn}\affiliation{\uw}
\author{V.E.~Guiseppe}\affiliation{\usc}
\author{C.R.~Haufe}\affiliation{\unc}\affiliation{\tunl}
\author{L.~Hehn}\affiliation{\lbnl}	
\author{R.~Henning}\affiliation{\unc}\affiliation{\tunl}
\author{E.W.~Hoppe}\affiliation{\pnnl}
\author{M.A.~Howe}\affiliation{\unc}\affiliation{\tunl}
\author{S.I.~Konovalov}\affiliation{\ITEP}
\author{R.T.~Kouzes}\affiliation{\pnnl}
\author{A.M.~Lopez}\affiliation{\ut}
\author{R.D.~Martin}\affiliation{\queens}
\author{R. Massarczyk}\email[Corresponding author : ]{massarczyk@lanl.gov}\affiliation{\lanl}
\author{S.J.~Meijer}\affiliation{\unc}\affiliation{\tunl}
\author{S.~Mertens}\affiliation{\mpi}\affiliation{\tum}	
\author{J.~Myslik}\affiliation{\lbnl}
\author{C. O'Shaughnessy}\altaffiliation{Present address: \lanl}\affiliation{\unc}\affiliation{\tunl}	
\author{G.~Othman}\affiliation{\unc}\affiliation{\tunl}
\author{W.~Pettus}\affiliation{\uw}
\author{A.W.P.~Poon}\affiliation{\lbnl}
\author{D.C.~Radford}\affiliation{\ornl}
\author{J.~Rager}\affiliation{\unc}\affiliation{\tunl}	
\author{A.L.~Reine}\affiliation{\unc}\affiliation{\tunl}
\author{K.~Rielage}\affiliation{\lanl}
\author{R.G.H.~Robertson}\affiliation{\uw}
\author{N.W.~Ruof}\affiliation{\uw}	
\author{B.~Shanks}\affiliation{\ornl}
\author{M.~Shirchenko}\affiliation{\JINR}
\author{A.M.~Suriano}\affiliation{\sdsmt}
\author{D.~Tedeschi}\affiliation{\usc}	
\author{R.L.~Varner}\affiliation{\ornl} 
\author{S. Vasilyev}\affiliation{\JINR}	
\author{K.~Vorren}\affiliation{\unc}\affiliation{\tunl} 
\author{B.R.~White}\affiliation{\lanl}
\author{J.F.~Wilkerson}\affiliation{\unc}\affiliation{\tunl}\affiliation{\ornl} \author{C. Wiseman}\affiliation{\usc}
\author{W.~Xu}\affiliation{\usd} 
\author{E.~Yakushev}\affiliation{\JINR}
\author{C.-H.~Yu}\affiliation{\ornl}
\author{V.~Yumatov}\affiliation{\ITEP}
\author{I.~Zhitnikov}\affiliation{\JINR} 
\author{B.X.~Zhu}\affiliation{\lanl}

\collaboration{{\sc{Majorana}} Collaboration}
\noaffiliation

\date{\today}

\begin{abstract}
The \MJ\ \DEM\ is an ultralow-background experiment searching for neutrinoless double-beta
decay in $^{76}$Ge. The heavily shielded array of germanium detectors, placed nearly a mile underground at the Sanford
Underground Research Facility in Lead, South Dakota, also allows searches for new exotic physics.
Free, relativistic, lightly ionizing particles with an electrical charge less than $e$ are forbidden by the standard model but predicted by some of its extensions. If such particles exist, they might be detected in the \MJ\ \DEM\ by searching for multiple-detector events with individual-detector energy depositions down to 1 keV. This search is background-free and no candidate events have been found in 285 days of data taking. New direct-detection limits are set for the flux of lightly ionizing particles for charges as low as $e/1000$.
\end{abstract}


\maketitle

Lightly ionizing particles (LIPs) are hypothetical particles for which the electromagnetic interaction is suppressed compared to particles like charged hadrons and leptons. A particle with a charge $q=e/f$ that is reduced by a factor $f$ relative to the electron charge $e$ is expected to have weaker electromagnetic interactions than standard singly charged particles. These particles are often referred to as milli- or minicharged particles (mCP) in the literature. In this work, we refer to them as LIPs, since this designation describes the energy loss phenomenology related to a class of detection techniques. The term LIPs includes mCPs since their signature would be diminished ionization, but it does not preclude other possible particles.

The standard model (SM) of particle physics does not include free fractionally charged particles \cite{Langacker1980} since the quarks are bound within hadrons and do not exist as free particles. However, the SM is known to be incomplete, since it cannot explain the nature of dark matter or dark energy. Unbound quarks, noninteger-charged bound states of quarks, or new leptons with fractional charge are a few possible candidates with LIP character that occur in proposed extensions of the SM. There are a variety of theories that permit an mCP including, for example, a fermion singlet~\cite{Okun1984,Vinyoles2016}, an additional mirror U(1) paraphoton that can mix with the photon~\cite{Holdom1985}, neutrinos with electromagnetic couplings~\cite{Foot1990}, vector particles~\cite{Gabrielli2015}, dark constituents bound to atoms~\cite{Cline2012}, charge quantization~\cite{Ignatiev1979,Wen1985,Schellekens1990,Babu1990}, or composite mCPs~\cite{Kouvaris2013}. The phenomenology of these models and their variants is very broad, justifying a variety of search techniques and leading to a rich experimental literature.

Although the masses of these particles can lie above the reach of current accelerators, experimental constraints on masses and charges of mCPs have been derived from fixed target accelerators~\cite{Aubert1983,Bergsma1984,Golowich1987,He1991,Huntrup1996,Ghosh1996,Prinz1998,Soper2014}, colliders~\cite{Abe1992,Buskulic1993,Akers1995,Abreu1997,Acosta2003,Abbiendi2003,Jaeckel2012,Chatrchyan2013}, stellar models \cite{Dobroliubov1990,Davidson2000,Feng2016,Vinyoles2016}, cosmic microwave background \cite{Ahlers2009,Burrage2009,Davidson2000,Dubovsky2003,Berezhiani2009,Dobroliubov1990,Vogel2013,Dolgov2017}, big-bang nucleosynthesis \cite{Davidson2000}, Supernova 1987A~\cite{Mohapatra1990,Davidson2000}, neutron stars~\cite{Huang2015,Korwar2017}, pulsars and gamma ray bursts~\cite{Kvam2014}, galaxy clusters~\cite{Kadota2016}, the Lamb shift~\cite{Dobroliubov1990,Davidson1991,Gluck2007}, dark cosmic ray searches~\cite{Hu2017}, positronium decay~\cite{Badertscher2006}, reactor neutrinos~\cite{Gninenko2006,Chen2014}, and the $\mu$ magnetic moment~\cite{Dobroliubov1990}. An early levitation experiment \cite{LaRue1981} found an indication for the existence for fractional charges that was not confirmed by following efforts~\cite{Smith1985, Moore2014}. Millikan's method is a long-standing technique to search for fractional charges~\cite{Kim2007}, combining the advantage of large probe sizes and high counting statistics. Brownian motion, however, limits this method's sensitivity \cite{Halyo2000}. Direct searches for LIPs, including MACRO~\cite{Ambrosio2000,Ambrosio2004}, Kamiokande-II~\cite{Mori1991}, and LSD~\cite{Aglietta1994} placed stringent limits on the LIP flux for $0.4 < f < 6$. The Cryogenic Dark Matter Search (CDMS) experiment~\cite{CDMS2010,Agnese2015} placed limits on exotic particles with $f < 200$ using a direct search technique. A 2009 review \cite{Perl2009} summarizes the experimental state of the field prior to the results of CDMS in 2010. References \cite{Vinyoles2016} and \cite{Haas2015} provide a broad list of references, give a recent overview of the results over the past decade, and discuss the mass-charge parameter space. Here, we describe an improved direct search for such particles.

\begin{figure}[t]
 \centering
 \includegraphics[width=0.80\columnwidth,keepaspectratio=true]{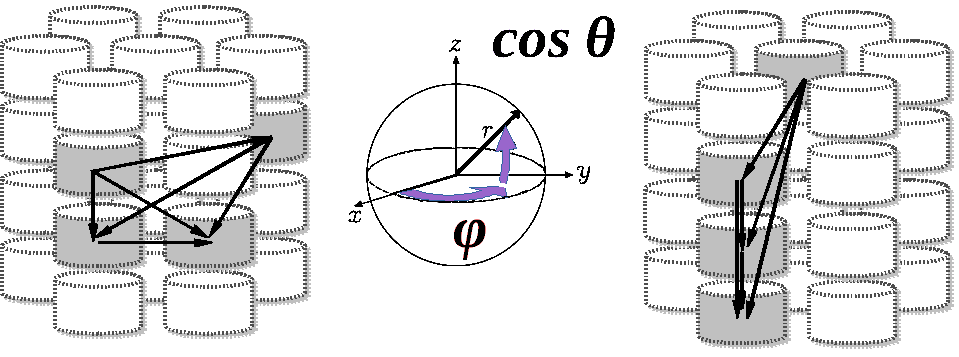}
 \caption{Sketch of the detector arrangement and the vectors used in background rejection cut. The grey shading indicates four detectors that triggered in this example. \textit{Left:} Vectors connecting the detector centers for a sample noise or background event, which do not point to a common location. \textit{Middle:} Definition of the angles used in the tracking algorithm. \textit{Right:} For a simulated LIP, the variation of directions ($\Delta$cos $\theta$ and $\Delta \,\phi$) is smaller.}
 \label{Fig_Directions}
\end{figure}

The \textsc{Majorana Demonstrator} \cite{Abgrall2014, Abgrall2017c} is located at a depth of 4850\,ft at the Sanford Underground
Research Facility \cite{Heise2015}. In addition to its primary goal of searching for neutrinoless double-beta decay, its ultralow-background configuration permits additional physics studies including searches for dark matter, axions, and exotic physics~\cite{Abgrall2017}. Two modules contain 44.1\,kg of high-purity germanium detectors, 29.7\,kg of which are enriched to 88\% $^{76}$Ge.
Fifty-eight detector units are installed in strings of three, four, or five detectors. The detector masses, diameters, and heights range from 0.5 to 1\,kg, 6 to 8\,cm, and 3to 4\,cm, respectively. A sketch of the setup can be seen in Fig.\,\ref{Fig_Directions} and a detailed description can be found in Ref. \cite{Abgrall2014}. The \MJ\ \DEM\ detectors are 3 to 5 times thicker than those used in CDMS, providing a higher sensitivity to lower-energy deposits per crossing and hence higher values of $f$ at comparable energy thresholds. The low thresholds, excellent energy resolution, reduced electronic noise, and pulse shape characteristics of the $p$-type point contact detectors ~\cite{luk89,Barbeau2007,Aguayo2011,Cooper2011} allow a competitive LIP search based on the \textsc{Demonstrator} data.

The analysis presented here includes data taken from June 2015 until March 2017. Excluding calibration, commissioning and blind data, the analyzed data include 285 days of live time, of which 121 days were taken with both modules operating in the final \DEM\ configuration~\cite{Abgrall2017c}. This corresponds to a total exposure of 4993 kg days for all detectors. Physics runs are typically one hour long. Since the set of operable detectors and their respective thresholds changed over the course of data taking, our simulation mirrored the changing conditions on a run-by-run basis. For several runs the threshold was increased to avoid noise introduced by external work during the construction phase. 

The flux ($\Phi (f)$) of LIPs through the detector array is given as:
\begin{equation}
 \Phi (f) = \frac{n}{\sum\limits_{i}\sum\limits_{m} A_{i,m} \epsilon_{i,m} t_{i} \Omega_{i,m}},
 \label{eq_1}
\end{equation}
where $n$ is the number of detected interactions. For zero candidates an upper bound on $\Phi$ can be set using the method of Feldman and Cousins in Ref. \cite{Feldman1998}. The sum index $i$ is over data runs and index $m$ is over the multiplicity values considered for LIP candidates. The multiplicity is defined as the number of detectors with signals above the threshold within a 4-$\mu$s-long coincidence window. We consider events with $m=$ 4, 5, and 6. The length of a run is given by its dead-time corrected live time ($t_i$). The detection efficiency ($\epsilon$) depends on each detector's threshold and the geometry of the active detectors, both of which vary run by run. On average, 70\% of the detectors are operable. The detection threshold was estimated by analyzing the baseline noise of each recorded waveform and verified in special forced trigger data. The detector baseline traces are processed with a trapezoidal filter. From the distribution of the integrated values of the flattop, we can estimate the energy at which we would detect events with a 99.7\% or greater probability. For the majority of the runs the individual thresholds are between 0.8 and 2\,keV.
The surface area ($A_{i,m}$) for an incident LIP is taken as the end cap area of the smallest detector crossed. For the \DEM\ detectors, $A=$30-37\,cm$^2$ ($\pm$1\,cm$^2$).
The \textsc{MaGe}~\cite{Bauer2006, Boswell2011} framework, based on \GF~\cite{Agostinelli2003} was used to estimate $A_{i,m}$ and the solid angle $\Omega$ for each run. Simulated noninteracting particles were used as a proxy for LIPs, and propagated through the the array with varying angles of incidence. Since the path length through detectors depends on the LIP trajectory angle through the array, the efficiency is a function of the incident angle therefore depends on the impinging flux distribution. CDMS assumed an isotropic distribution from above \cite{Agnese2015}. We present results for that same distribution for comparison as well as results for a cos$^2\theta$ distribution, where $\theta$ is the polar angle. The latter function is a proxy for particles created in the upper atmosphere \cite{Perl2009}.
For $m=4$ events, the average solid angle is $\sim$2.4\,sr (1.5\,sr) for a uniform flux from above (cos$^2\theta$ distribution). The exact number varies for each run. Larger $m$'s have a smaller number of possible detector combinations and, hence smaller $\Omega$. For $m=5$ and 6, the average solid angles are 1\,sr (0.6\,sr) and 0.06\,sr (0.02\,sr), respectively.

\begin{figure}[]
 \centering
 \includegraphics[width=0.90\columnwidth,keepaspectratio=true]{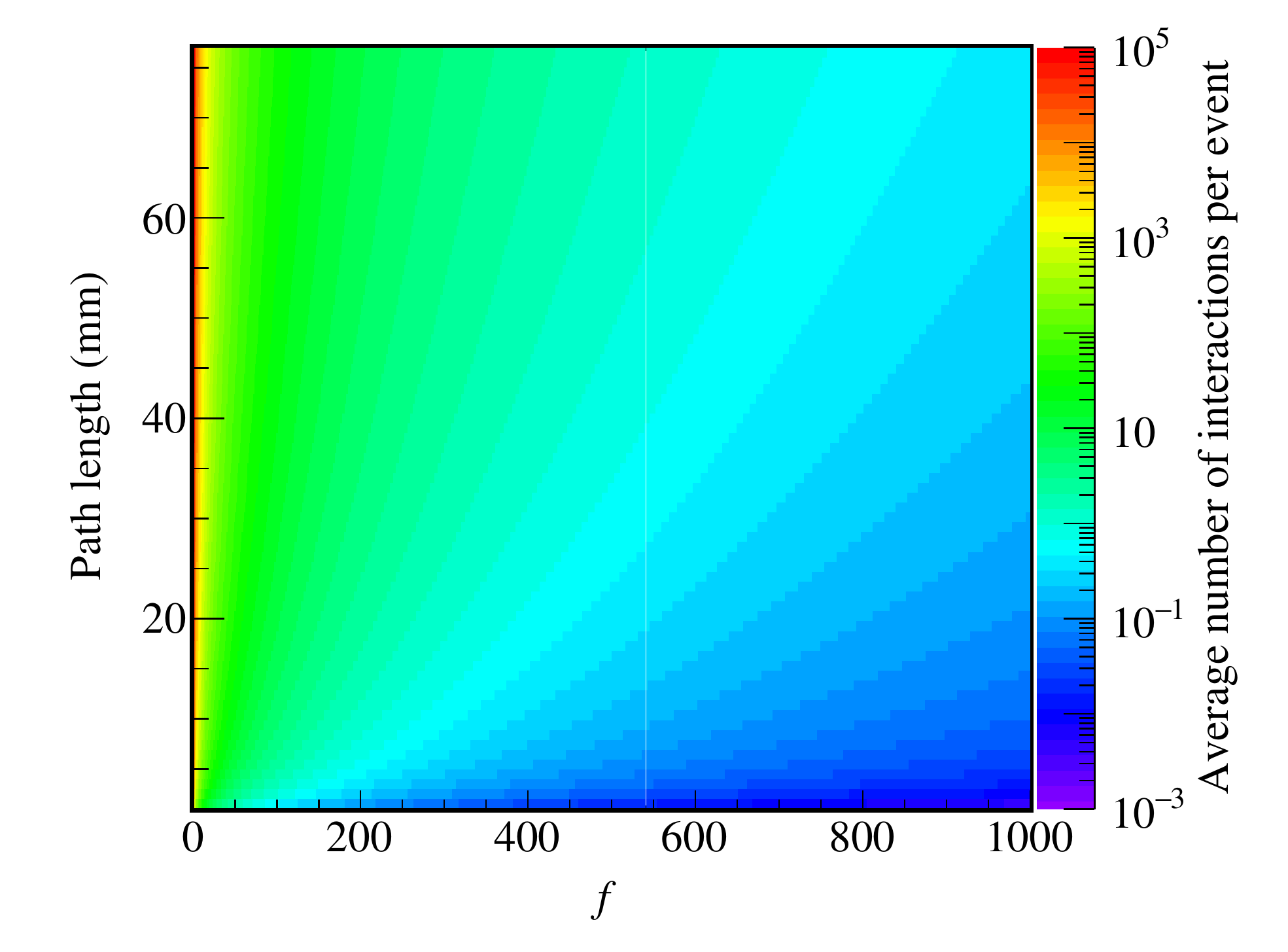}
 \caption{Average number of interactions for one LIP event in germanium as a function of path length and the parameter $f = e/q$.}
 \label{Fig_Number}
\end{figure}

\begin{figure}[h]
 \centering
 \includegraphics[width=0.90\columnwidth,keepaspectratio=true]{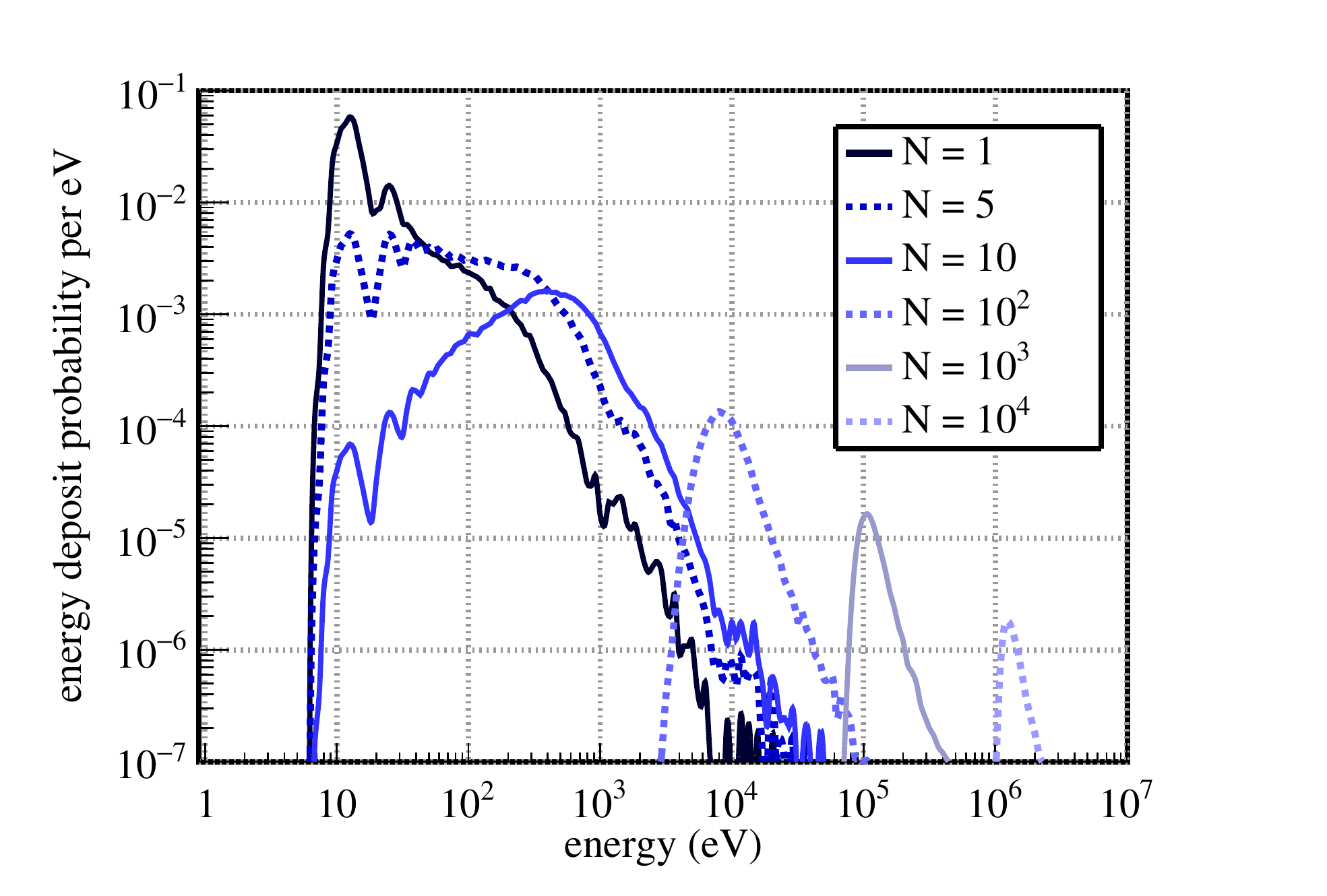}
 \caption{Expected energy loss for several numbers of interaction $N$. All curves are calculated Poisson-weighted convolutions of the single interaction distribution.}
 \label{Fig_CollisionEnergy}
\end{figure}

\begin{figure}[]
 \centering
 \includegraphics[width=0.90\columnwidth,keepaspectratio=true]{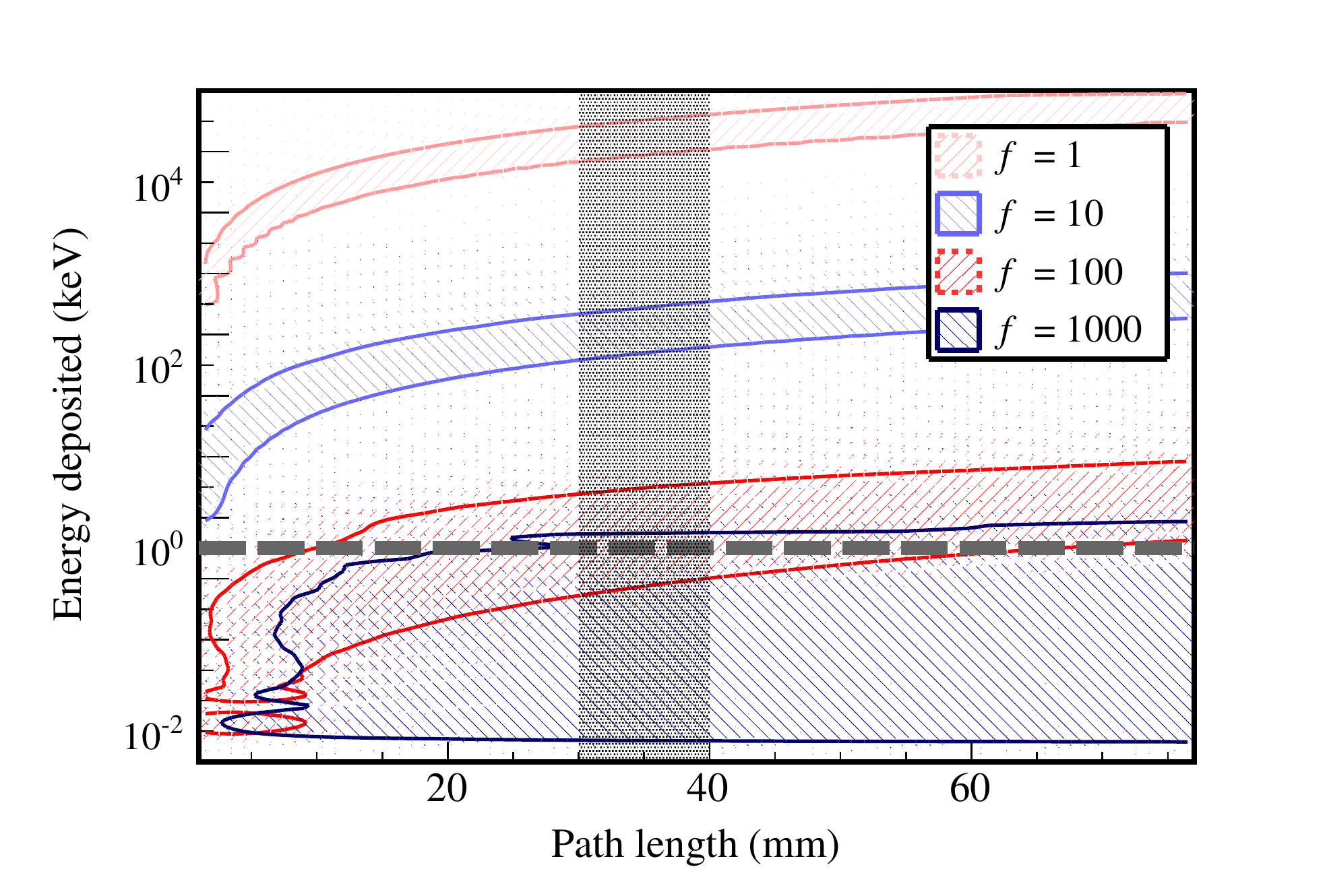}
 \caption{Energy depositions of simulated events with different path lengths in germanium. The colored bands indicate the 90\% enclosure of all events for four different values of the denominator $f$, respectively. The horizontal dark gray dashed line indicates a 1\,keV threshold, the vertical gray area indicates the average detector thickness.}
 \label{Fig_Length}
\end{figure}

For large $f$, LIPs interact potentially only once in a detector, cf. Fig.\,\ref{Fig_Number}, leading to large energy-deposit fluctuations. Following Refs. ~\cite{Agnese2015,Prasad2013}, we calculate the expected energy-loss distribution based on the single-interaction energy loss. The photoabsorption ionization model~\cite{Allison1980} was used to calculate the interaction cross section.
This probability distribution function (PDF) for the single-interaction energy loss is convolved with itself $N$ times to derive the PDF for $N$ such interactions~\cite{Bichsel2006}, cf. Fig.\,\ref{Fig_CollisionEnergy}. The number of interactions per unit path length through a detector was calculated using the approach of Ref.~\cite{Bichsel1988}. The result is a function of $f$ as shown in Fig.~\ref{Fig_Number}. The expected energy deposited as a function of track length and $f$ is shown in Fig.~\ref{Fig_Length}. The probability that a LIP with $f$ deposits enough energy to exceed the detector threshold is calculated for simulated events. The total efficiency is the product of these individual detector probabilities.

For each run and detector, the data acquisition threshold is applied in combination with the simulation, resulting in a run-dependent detection efficiency for LIPs with a given $m$ and trajectory. The simulated efficiency distributions also take into account inoperable channels and exclude them from the analysis.

\begin{figure}[t]
 \centering
 \includegraphics[width=0.90\columnwidth,keepaspectratio=true]{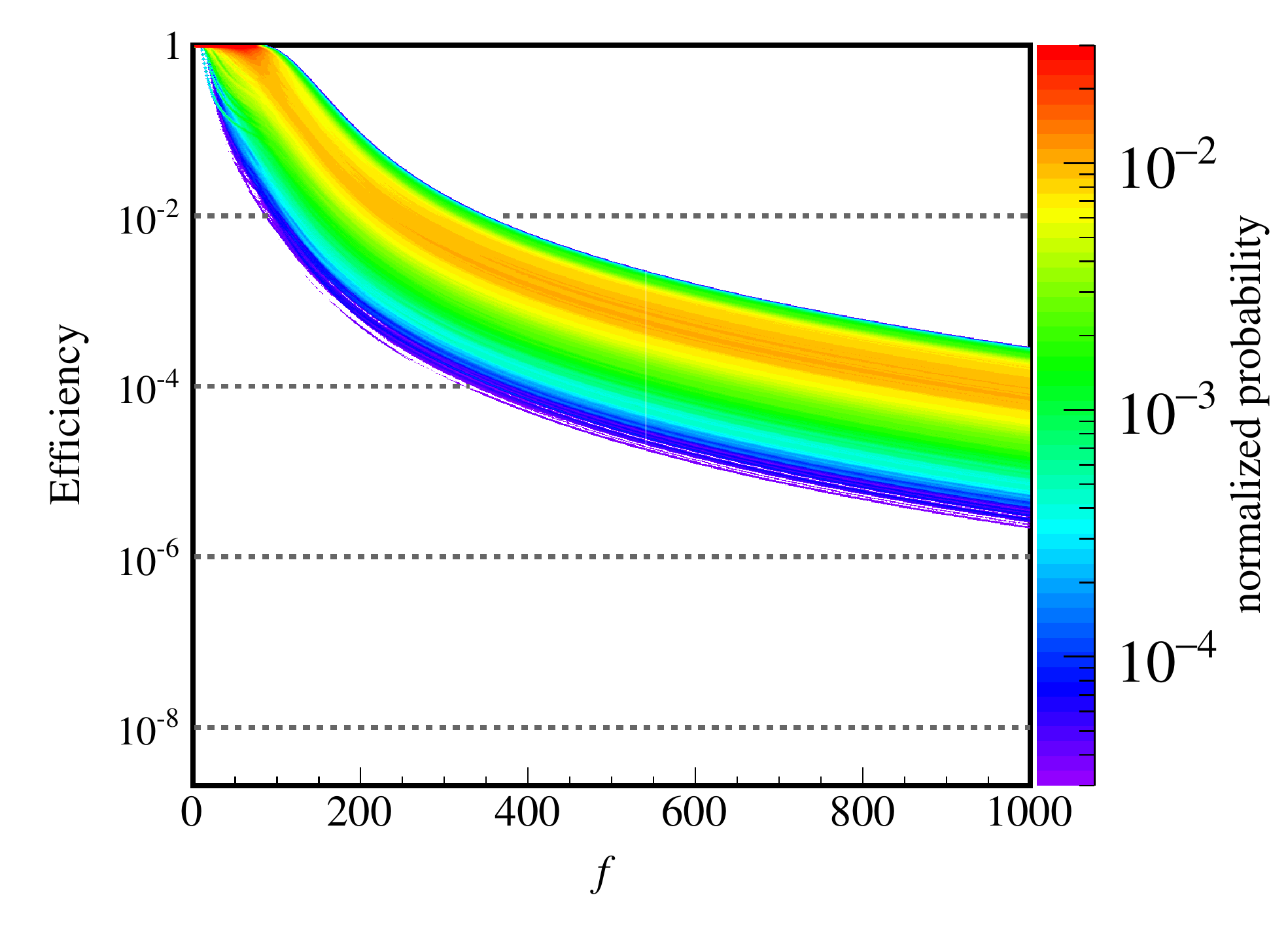}
 \caption{Detection efficiency for $m=4$ events for one data set as a function of $f$, the factor by which the charge of the particle is smaller than the elementary charge. For each individual $f$ the probabilities of the detection efficiencies are given by the color scale.}
 \label{Fig_Efficiency}
\end{figure}

Two factors give non-negligible contributions to the uncertainty of the efficiency $\epsilon$. One is the uncertainty in the traversed-detector path length that determines the number of interactions. This results from uncertainty in the thickness of the dead layer at the outer surface of each detector. The other factor is that the detectors have a finite energy resolution. Both effects contribute to the uncertainty in the probability that a LIP energy deposit will be above the  threshold, especially for large $f$ and small energy depositions. In order to estimate the systematic uncertainties, we analyzed the simulated efficiencies 100 times for each individual run, varying the track length $l$ inside each detector traversed, and the energy resolution. The values were drawn from Gaussian distributions around the mean value of each parameter, with widths $\sigma_l = \pm$1\,mm, $\sigma_n = \sqrt{n}$, and FWHM$_\mathrm{E}$ = 0.2\,keV, respectively. The energy resolution value corresponds to the FWHM below 10\,keV in the \DEM\,~\cite{Abgrall2017}. Finally all the efficiencies for a given data set and multiplicity $m$ are combined in one histogram. In Fig.~\ref{Fig_Efficiency}, the distribution of efficiencies for $m=4$ events is drawn. The width of the distribution for each value of $f$ is used as the systematic uncertainty. This conservative approach allows us to show that our sensitivity is mostly independent of short-lived variations in detector settings.

\begin{figure}[t]
 \centering
 \includegraphics[width=0.90\columnwidth,keepaspectratio=true]{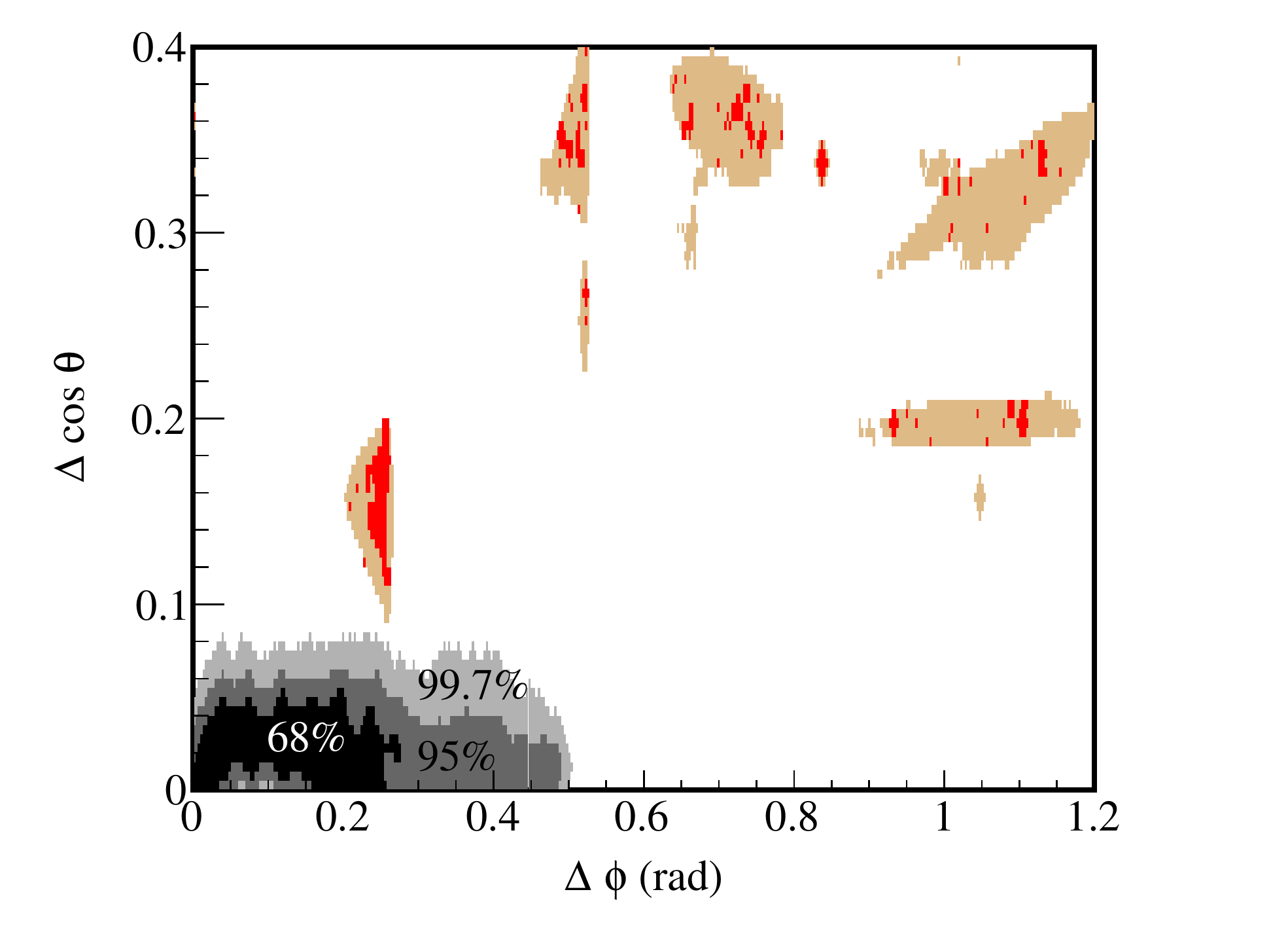}
 \caption{Event-by-event spread in the two spherical angles, see Fig.~\ref{Fig_Directions}, for \DEM\ data (brown) during the 285-day lifetime period. Events within a one-second coincidence with the muon veto~\cite{Abgrall2017b} are shown in red. Almost 25\,000 events survive the basic multiplicity cut applied for this study. For simulated LIPs from an isotropic source, the regions that include 68\,\% (black), 95\,\% (dark gray), and 99.7\,\% (used in analysis, light gray) of all events are shown. }
 \label{Fig_Cut}
\end{figure}

 In each detector within the \DEM, a fair number of nonphysics events contribute to the low-energy backgrounds. These include noise, microphonics during nitrogen fills, and pulser cross talk. A multiplicity requirement of $4\leq m\leq 6$ eliminates the majority ($\approx$97\%) of these without significant additional analysis. In addition, a one-second anticoincidence time with the muon veto of the \textsc{Demonstrator} excludes cosmogenic background. All events surviving events from the 285 days of live time (corresponding to 4993 kg days) are depicted in Fig.~\ref{Fig_Cut}. There is no requirement on the geometric arrangement allowing us to greatly increase $\Omega$, and therefore sensitivity, relative to the CDMS experiment. In other words, instead of searching only for particles from above, we also search for LIPs that traverse multiple strings.

Because of the variation of detector sizes and variety of possible LIP trajectories, it is impossible to include a CDMS-like energy consistency requirement; the path lengths in different detectors are not necessarily comparable. To reduce the remaining background within the high-$m$ sample, a tracking algorithm was applied. Each candidate event is compared to the simulated signature of a LIP. A LIP will traverse the array in a straight line and vectors connecting pairs of triggered detectors (see the rightmost panel of Fig.~\ref{Fig_Directions}) should all point roughly to the same direction on an imaginary sphere surrounding the array. Since the exact location of the interaction within the detector is unknown, the center of the detectors is used as start and end point of each vector. The direction of these vectors can be described with two angles using spherical coordinates $\theta$ and $\phi$, depicted in Fig.~\ref{Fig_Directions}.

Since their triggered detectors do not fall along a single track, events due to instrumental effects and internal backgrounds will self-evidently have larger values of $\Delta\theta$ and $\Delta\phi$, the differences in $\theta$ and $\phi$ determined from different detector pairings in a single event. Distinguishing muons from LIPs with the tracking algorithm may seem more difficult; we can study such tracks by choosing events that are triggered in coincidence with the muon veto system. A minimally ionizing LIP with high $f$ ($>6$) would not deposit enough energy to trigger the veto, which is made of 2-inch-thick plastic scintillator panels and has a trigger threshold of 1\,MeV. For muon events, the particle shower accompanying the muon tends to trigger more than six detectors, or additional out-of-line detectors, as shown by the red muon veto-coincident events in Fig.~\ref{Fig_Cut}. Simulations show that LIPs with $f>6$ do not produce significant showers, unlike muons. For $f$=1, 90\% of the events are accompanied by a shower. For $f$=6 this number drops to only 7\% and can be assumed to be close to zero for higher $f$. Therefore, the simulated LIP events show smaller spreads in $\theta$ and $\phi$ values than muon events. Since our analysis requires linelike shower-free events we excluded limits below $f=6$ from our results.

A cut in $\Delta\theta$ and $\Delta\phi$, shown by the gray region in Fig.~\ref{Fig_Cut}, was chosen based on the LIP simulations. The efficiency for retaining a LIP candidate in the tracking algorithm is effectively unity with an uncertainty of less than 0.3\%, which is negligible compared to the other uncertainties. Restricting the multiplicities to $m=4$, 5, or 6 events in the \DEM\ data, we find no LIP candidate events in the shaded area. Applying the Feldman and Cousins procedure \cite{Feldman1998}, a value of 2.44 (90\% C.L.) is used as the upper limit for $n$ in Eq.\,(\ref{eq_1}).

\begin{figure}[t]
 \centering
 \includegraphics[width=0.90\columnwidth,keepaspectratio=true]{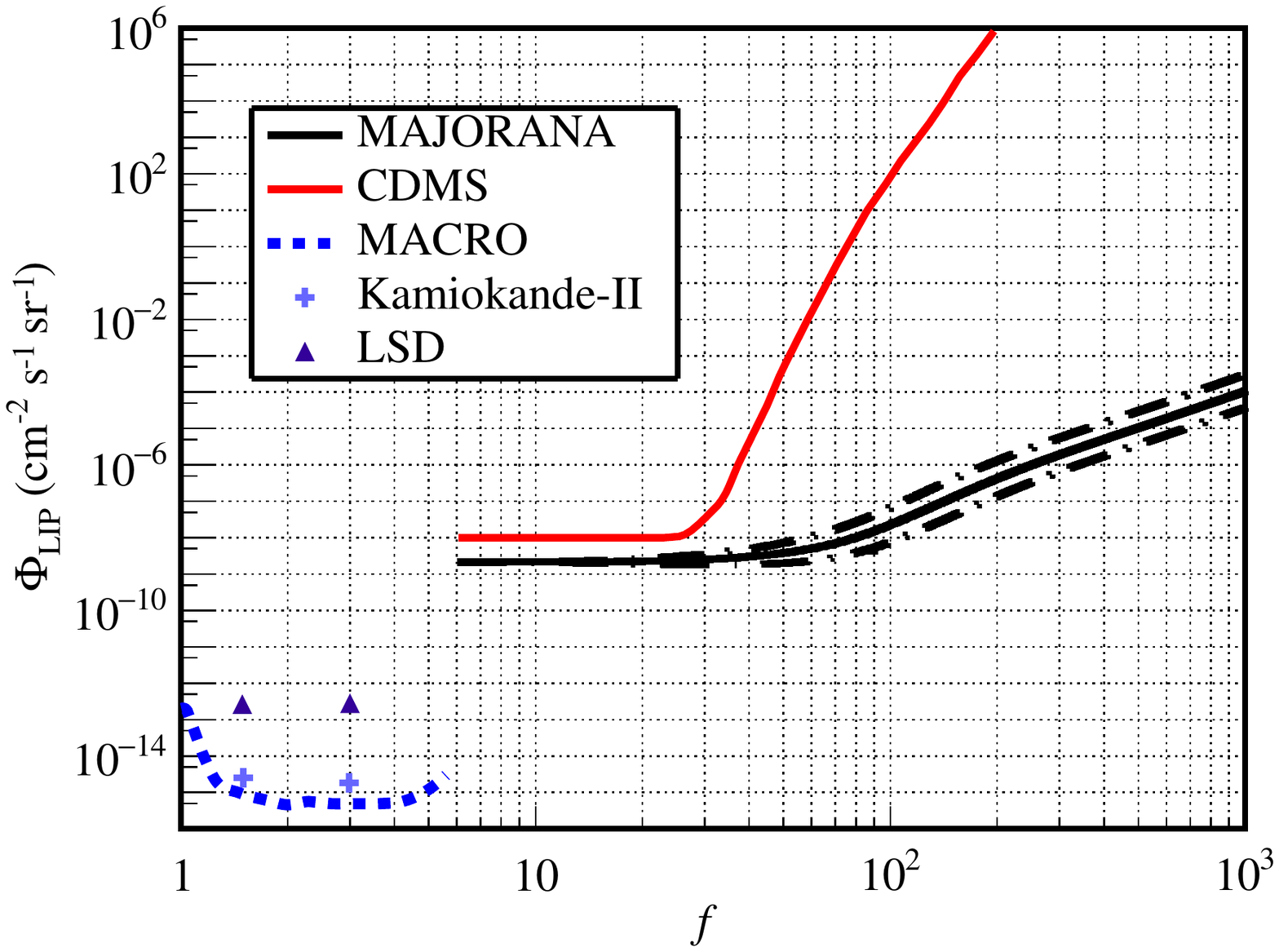}
 \caption{LIP flux limit from above on the \textsc{Majorana Demonstrator} using a 90\% confidence level (black line) and its 1-$\sigma$ uncertainty bands (dashed black lines). Results from MACRO \cite{Ambrosio2004}, Kamiokande-II \cite{Mori1991}, LSD \cite{Aglietta1994}, and CDMS \cite{Agnese2015} are shown as well. All limits assume an isotropic flux. As indicated in the text, a cos$^2\theta$ distribution of LIPs would result in a 38\% less restrictive curve. }
 \label{Fig_Result}
\end{figure}

Figure~\ref{Fig_Result} displays the results as a function of $f$. For charges between $e/6$ and $e/30$, a limit of 2$\times$10$^{-9}$ particles per cm$^{2}$\,s\,sr is found. A deviation from the minimally ionizing character ($\beta\gamma\sim3$) of the particle would result in a higher detection efficiency. Hence, the limits presented are conservative upper limits. Using the assumption that LIPs are impinging with a cos$^2\theta$ distribution would result in a slightly smaller detection efficiency and, therefore, in a limit that is about 38\% above that of the isotropic model. 

This work presents the first limits on massive relativistic particles with a fractional charge using the unique features of the \MJ\ \DEM. The large path length due to thick detectors in combination with the low thresholds allows for sensitivity down to 1/1000$^{th}$ of an elementary charge. These are the first results for a nonaccelerator based experiment on the natural flux of lightly ionizing particles with charges less than $e$/200 and an improvement of the existing limits for charges between $e$/6 and $e$/200. The results presented will help to exclude certain models or at least restrict their parameter space, e.g. for the millicharged dark matter presented in Ref.\cite{Hu2017}.
\begin{acknowledgements}
This material is based upon work supported by the U.S. Department of Energy, Office of Science, Office of Nuclear Physics under Award
Numbers DE-AC02-05CH11231, DE-AC05-00OR22725, DE-AC05-76RL0130, DE-AC52-06NA25396, DE-FG02-97ER41020, DE-FG02-97ER41033, DE-FG02-97ER41041, DE-SC0010254, DE-SC0012612, DE-SC0014445, and DE-SC0018060. We acknowledge support from the Particle Astrophysics Program and Nuclear Physics Program of the National Science Foundation
through grant numbers MRI-0923142, PHY-1003399, PHY-1102292, PHY-1206314, and PHY-1614611. We gratefully acknowledge the support of the U.S. Department of Energy through the LANL/LDRD Program and through the PNNL/LDRD Program for this work. We acknowledge support from the Russian Foundation for Basic Research, grant No. 15-02-02919.
We acknowledge the support of the Natural Sciences and Engineering Research Council of Canada, funding reference number SAPIN-2017-00023, and from the Canada Foundation from Innovation John R. Evans Leaders Fund. We thank the Yamaha Science Foundation Japan for their support.
This research used resources provided by the Oak Ridge Leadership Computing Facility at Oak Ridge National Laboratory and by
the National Energy Research Scientific Computing Center, a DOE Office of Science User Facility. We thank our hosts and colleagues at the Sanford Underground Research Facility for their support.
\end{acknowledgements}
\bibliography{lips}

\begin{thebibliography}{78}%
\makeatletter
\providecommand \@ifxundefined [1]{%
 \@ifx{#1\undefined}
}%
\providecommand \@ifnum [1]{%
 \ifnum #1\expandafter \@firstoftwo
 \else \expandafter \@secondoftwo
 \fi
}%
\providecommand \@ifx [1]{%
 \ifx #1\expandafter \@firstoftwo
 \else \expandafter \@secondoftwo
 \fi
}%
\providecommand \natexlab [1]{#1}%
\providecommand \enquote  [1]{``#1''}%
\providecommand \bibnamefont  [1]{#1}%
\providecommand \bibfnamefont [1]{#1}%
\providecommand \citenamefont [1]{#1}%
\providecommand \href@noop [0]{\@secondoftwo}%
\providecommand \href [0]{\begingroup \@sanitize@url \@href}%
\providecommand \@href[1]{\@@startlink{#1}\@@href}%
\providecommand \@@href[1]{\endgroup#1\@@endlink}%
\providecommand \@sanitize@url [0]{\catcode `\\12\catcode `\$12\catcode
  `\&12\catcode `\#12\catcode `\^12\catcode `\_12\catcode `\%12\relax}%
\providecommand \@@startlink[1]{}%
\providecommand \@@endlink[0]{}%
\providecommand \url  [0]{\begingroup\@sanitize@url \@url }%
\providecommand \@url [1]{\endgroup\@href {#1}{\urlprefix }}%
\providecommand \urlprefix  [0]{URL }%
\providecommand \Eprint [0]{\href }%
\providecommand \doibase [0]{http://dx.doi.org/}%
\providecommand \selectlanguage [0]{\@gobble}%
\providecommand \bibinfo  [0]{\@secondoftwo}%
\providecommand \bibfield  [0]{\@secondoftwo}%
\providecommand \translation [1]{[#1]}%
\providecommand \BibitemOpen [0]{}%
\providecommand \bibitemStop [0]{}%
\providecommand \bibitemNoStop [0]{.\EOS\space}%
\providecommand \EOS [0]{\spacefactor3000\relax}%
\providecommand \BibitemShut  [1]{\csname bibitem#1\endcsname}%
\let\auto@bib@innerbib\@empty
\bibitem [{\citenamefont {Langacker}\ and\ \citenamefont
  {Pi}(1980)}]{Langacker1980}%
  \BibitemOpen
  \bibfield  {author} {\bibinfo {author} {\bibfnamefont {P.}~\bibnamefont
  {Langacker}}\ and\ \bibinfo {author} {\bibfnamefont {S.-Y.}\ \bibnamefont
  {Pi}},\ }\href {\doibase 10.1103/PhysRevLett.45.1} {\bibfield  {journal}
  {\bibinfo  {journal} {Phys. Rev. Lett.}\ }\textbf {\bibinfo {volume} {45}},\
  \bibinfo {pages} {1} (\bibinfo {year} {1980})}\BibitemShut {NoStop}%
\bibitem [{\citenamefont {Okun}\ \emph {et~al.}(1984)\citenamefont {Okun},
  \citenamefont {Voloshin},\ and\ \citenamefont {Zakharov}}]{Okun1984}%
  \BibitemOpen
  \bibfield  {author} {\bibinfo {author} {\bibfnamefont {L.}~\bibnamefont
  {Okun}}, \bibinfo {author} {\bibfnamefont {M.}~\bibnamefont {Voloshin}}, \
  and\ \bibinfo {author} {\bibfnamefont {V.}~\bibnamefont {Zakharov}},\ }\href
  {\doibase https://doi.org/10.1016/0370-2693(84)91884-7} {\bibfield  {journal}
  {\bibinfo  {journal} {Physics Letters B}\ }\textbf {\bibinfo {volume}
  {138}},\ \bibinfo {pages} {115 } (\bibinfo {year} {1984})}\BibitemShut
  {NoStop}%
\bibitem [{\citenamefont {Vinyoles}\ and\ \citenamefont
  {Vogel}(2016)}]{Vinyoles2016}%
  \BibitemOpen
  \bibfield  {author} {\bibinfo {author} {\bibfnamefont {N.}~\bibnamefont
  {Vinyoles}}\ and\ \bibinfo {author} {\bibfnamefont {H.}~\bibnamefont
  {Vogel}},\ }\href {http://stacks.iop.org/1475-7516/2016/i=03/a=002}
  {\bibfield  {journal} {\bibinfo  {journal} {Journal of Cosmology and
  Astroparticle Physics}\ }\textbf {\bibinfo {volume} {2016}},\ \bibinfo
  {pages} {002} (\bibinfo {year} {2016})}\BibitemShut {NoStop}%
\bibitem [{\citenamefont {Holdom}(1986)}]{Holdom1985}%
  \BibitemOpen
  \bibfield  {author} {\bibinfo {author} {\bibfnamefont {B.}~\bibnamefont
  {Holdom}},\ }\href {\doibase 10.1016/0370-2693(86)91377-8} {\bibfield
  {journal} {\bibinfo  {journal} {Phys. Lett.}\ }\textbf {\bibinfo {volume}
  {166B}},\ \bibinfo {pages} {196} (\bibinfo {year} {1986})}\BibitemShut
  {NoStop}%
\bibitem [{\citenamefont {Foot}\ \emph {et~al.}(1990)\citenamefont {Foot},
  \citenamefont {Joshi}, \citenamefont {Lew},\ and\ \citenamefont
  {Volkas}}]{Foot1990}%
  \BibitemOpen
  \bibfield  {author} {\bibinfo {author} {\bibfnamefont {R.}~\bibnamefont
  {Foot}}, \bibinfo {author} {\bibfnamefont {G.}~\bibnamefont {Joshi}},
  \bibinfo {author} {\bibfnamefont {H.}~\bibnamefont {Lew}}, \ and\ \bibinfo
  {author} {\bibfnamefont {R.}~\bibnamefont {Volkas}},\ }\href@noop {}
  {\bibfield  {journal} {\bibinfo  {journal} {Mod. Phys. Lett. A}\ }\textbf
  {\bibinfo {volume} {5}},\ \bibinfo {pages} {95} (\bibinfo {year}
  {1990})}\BibitemShut {NoStop}%
\bibitem [{\citenamefont {Gabrielli}\ \emph {et~al.}(2015)\citenamefont
  {Gabrielli}, \citenamefont {Marzola}, \citenamefont {Raidal},\ and\
  \citenamefont {Veerm{\"a}e}}]{Gabrielli2015}%
  \BibitemOpen
  \bibfield  {author} {\bibinfo {author} {\bibfnamefont {E.}~\bibnamefont
  {Gabrielli}}, \bibinfo {author} {\bibfnamefont {L.}~\bibnamefont {Marzola}},
  \bibinfo {author} {\bibfnamefont {M.}~\bibnamefont {Raidal}}, \ and\ \bibinfo
  {author} {\bibfnamefont {H.}~\bibnamefont {Veerm{\"a}e}},\ }\href {\doibase
  10.1007/JHEP08(2015)150} {\bibfield  {journal} {\bibinfo  {journal} {Journal
  of High Energy Physics}\ }\textbf {\bibinfo {volume} {2015}},\ \bibinfo
  {pages} {150} (\bibinfo {year} {2015})}\BibitemShut {NoStop}%
\bibitem [{\citenamefont {Cline}\ \emph {et~al.}(2012)\citenamefont {Cline},
  \citenamefont {Liu},\ and\ \citenamefont {Xue}}]{Cline2012}%
  \BibitemOpen
  \bibfield  {author} {\bibinfo {author} {\bibfnamefont {J.~M.}\ \bibnamefont
  {Cline}}, \bibinfo {author} {\bibfnamefont {Z.}~\bibnamefont {Liu}}, \ and\
  \bibinfo {author} {\bibfnamefont {W.}~\bibnamefont {Xue}},\ }\href
  {arXiv:1201.4858} {\bibfield  {journal} {\bibinfo  {journal} {Phys. Rev. D}\
  }\textbf {\bibinfo {volume} {85}},\ \bibinfo {pages} {101302} (\bibinfo
  {year} {2012})}\BibitemShut {NoStop}%
\bibitem [{\citenamefont {Ignatiev}\ \emph {et~al.}(1979)\citenamefont
  {Ignatiev}, \citenamefont {Kuzmin},\ and\ \citenamefont
  {Shaposhnikov}}]{Ignatiev1979}%
  \BibitemOpen
  \bibfield  {author} {\bibinfo {author} {\bibfnamefont {A.~Y.}\ \bibnamefont
  {Ignatiev}}, \bibinfo {author} {\bibfnamefont {V.~A.}\ \bibnamefont
  {Kuzmin}}, \ and\ \bibinfo {author} {\bibfnamefont {M.~E.}\ \bibnamefont
  {Shaposhnikov}},\ }\href@noop {} {\bibfield  {journal} {\bibinfo  {journal}
  {Phys. Lett., B}\ }\textbf {\bibinfo {volume} {84}},\ \bibinfo {pages} {315}
  (\bibinfo {year} {1979})}\BibitemShut {NoStop}%
\bibitem [{\citenamefont {Wen}\ and\ \citenamefont {Witten}(1985)}]{Wen1985}%
  \BibitemOpen
  \bibfield  {author} {\bibinfo {author} {\bibfnamefont {X.-G.}\ \bibnamefont
  {Wen}}\ and\ \bibinfo {author} {\bibfnamefont {E.}~\bibnamefont {Witten}},\
  }\href {\doibase https://doi.org/10.1016/0550-3213(85)90592-9} {\bibfield
  {journal} {\bibinfo  {journal} {Nuclear Physics B}\ }\textbf {\bibinfo
  {volume} {261}},\ \bibinfo {pages} {651 } (\bibinfo {year}
  {1985})}\BibitemShut {NoStop}%
\bibitem [{\citenamefont {Schellekens}(1990)}]{Schellekens1990}%
  \BibitemOpen
  \bibfield  {author} {\bibinfo {author} {\bibfnamefont {A.}~\bibnamefont
  {Schellekens}},\ }\href {\doibase
  https://doi.org/10.1016/0370-2693(90)91190-M} {\bibfield  {journal} {\bibinfo
   {journal} {Physics Letters B}\ }\textbf {\bibinfo {volume} {237}},\ \bibinfo
  {pages} {363 } (\bibinfo {year} {1990})}\BibitemShut {NoStop}%
\bibitem [{\citenamefont {Babu}\ and\ \citenamefont
  {Mohapatra}(1990)}]{Babu1990}%
  \BibitemOpen
  \bibfield  {author} {\bibinfo {author} {\bibfnamefont {K.~S.}\ \bibnamefont
  {Babu}}\ and\ \bibinfo {author} {\bibfnamefont {R.~N.}\ \bibnamefont
  {Mohapatra}},\ }\href {\doibase 10.1103/PhysRevD.41.271} {\bibfield
  {journal} {\bibinfo  {journal} {Phys. Rev. D}\ }\textbf {\bibinfo {volume}
  {41}},\ \bibinfo {pages} {271} (\bibinfo {year} {1990})}\BibitemShut
  {NoStop}%
\bibitem [{\citenamefont {Kouvaris}(2013)}]{Kouvaris2013}%
  \BibitemOpen
  \bibfield  {author} {\bibinfo {author} {\bibfnamefont {C.}~\bibnamefont
  {Kouvaris}},\ }\href {\doibase 10.1103/PhysRevD.88.015001} {\bibfield
  {journal} {\bibinfo  {journal} {Phys. Rev. D}\ }\textbf {\bibinfo {volume}
  {88}},\ \bibinfo {pages} {015001} (\bibinfo {year} {2013})}\BibitemShut
  {NoStop}%
\bibitem [{\citenamefont {Aubert}\ \emph {et~al.}(1983)\citenamefont {Aubert}
  \emph {et~al.}}]{Aubert1983}%
  \BibitemOpen
  \bibfield  {author} {\bibinfo {author} {\bibfnamefont {J.}~\bibnamefont
  {Aubert}} \emph {et~al.},\ }\href@noop {} {\bibfield  {journal} {\bibinfo
  {journal} {Phys. Lett. B}\ }\textbf {\bibinfo {volume} {133}},\ \bibinfo
  {pages} {461} (\bibinfo {year} {1983})}\BibitemShut {NoStop}%
\bibitem [{\citenamefont {Bergsma}\ \emph {et~al.}(1984)\citenamefont {Bergsma}
  \emph {et~al.}}]{Bergsma1984}%
  \BibitemOpen
  \bibfield  {author} {\bibinfo {author} {\bibfnamefont {F.}~\bibnamefont
  {Bergsma}} \emph {et~al.},\ }\href@noop {} {\bibfield  {journal} {\bibinfo
  {journal} {Z. Phys. C}\ }\textbf {\bibinfo {volume} {24}},\ \bibinfo {pages}
  {217} (\bibinfo {year} {1984})}\BibitemShut {NoStop}%
\bibitem [{\citenamefont {Golowich}\ and\ \citenamefont
  {Robinett}(1987)}]{Golowich1987}%
  \BibitemOpen
  \bibfield  {author} {\bibinfo {author} {\bibfnamefont {E.}~\bibnamefont
  {Golowich}}\ and\ \bibinfo {author} {\bibfnamefont {R.~W.}\ \bibnamefont
  {Robinett}},\ }\href {\doibase 10.1103/PhysRevD.35.391} {\bibfield  {journal}
  {\bibinfo  {journal} {Phys. Rev. D}\ }\textbf {\bibinfo {volume} {35}},\
  \bibinfo {pages} {391} (\bibinfo {year} {1987})}\BibitemShut {NoStop}%
\bibitem [{\citenamefont {He}\ and\ \citenamefont {Price}(1991)}]{He1991}%
  \BibitemOpen
  \bibfield  {author} {\bibinfo {author} {\bibfnamefont {Y.~D.}\ \bibnamefont
  {He}}\ and\ \bibinfo {author} {\bibfnamefont {P.~B.}\ \bibnamefont {Price}},\
  }\href@noop {} {\bibfield  {journal} {\bibinfo  {journal} {Phys. Rev. C}\
  }\textbf {\bibinfo {volume} {44}},\ \bibinfo {pages} {1672} (\bibinfo {year}
  {1991})}\BibitemShut {NoStop}%
\bibitem [{\citenamefont {H\"untrup}\ \emph {et~al.}(1996)\citenamefont
  {H\"untrup}, \citenamefont {Weidmann}, \citenamefont {Hirzebruch},
  \citenamefont {Winkel},\ and\ \citenamefont {Heinrich}}]{Huntrup1996}%
  \BibitemOpen
  \bibfield  {author} {\bibinfo {author} {\bibfnamefont {G.}~\bibnamefont
  {H\"untrup}}, \bibinfo {author} {\bibfnamefont {D.}~\bibnamefont {Weidmann}},
  \bibinfo {author} {\bibfnamefont {S.~E.}\ \bibnamefont {Hirzebruch}},
  \bibinfo {author} {\bibfnamefont {E.}~\bibnamefont {Winkel}}, \ and\ \bibinfo
  {author} {\bibfnamefont {W.}~\bibnamefont {Heinrich}},\ }\href {\doibase
  10.1103/PhysRevC.53.358} {\bibfield  {journal} {\bibinfo  {journal} {Phys.
  Rev. C}\ }\textbf {\bibinfo {volume} {53}},\ \bibinfo {pages} {358} (\bibinfo
  {year} {1996})}\BibitemShut {NoStop}%
\bibitem [{\citenamefont {Ghosh}\ \emph {et~al.}(1996)\citenamefont {Ghosh}
  \emph {et~al.}}]{Ghosh1996}%
  \BibitemOpen
  \bibfield  {author} {\bibinfo {author} {\bibfnamefont {D.}~\bibnamefont
  {Ghosh}} \emph {et~al.},\ }\href@noop {} {\bibfield  {journal} {\bibinfo
  {journal} {Fizika B}\ }\textbf {\bibinfo {volume} {5}},\ \bibinfo {pages}
  {135} (\bibinfo {year} {1996})}\BibitemShut {NoStop}%
\bibitem [{\citenamefont {Prinz}\ \emph {et~al.}(1998)\citenamefont {Prinz}
  \emph {et~al.}}]{Prinz1998}%
  \BibitemOpen
  \bibfield  {author} {\bibinfo {author} {\bibfnamefont {A.~A.}\ \bibnamefont
  {Prinz}} \emph {et~al.},\ }\href {\doibase 10.1103/PhysRevLett.81.1175}
  {\bibfield  {journal} {\bibinfo  {journal} {Phys. Rev. Lett.}\ }\textbf
  {\bibinfo {volume} {81}},\ \bibinfo {pages} {1175} (\bibinfo {year}
  {1998})}\BibitemShut {NoStop}%
\bibitem [{\citenamefont {Soper}\ \emph {et~al.}(2014)\citenamefont {Soper},
  \citenamefont {Spannowsky}, \citenamefont {Wallace},\ and\ \citenamefont
  {Tait}}]{Soper2014}%
  \BibitemOpen
  \bibfield  {author} {\bibinfo {author} {\bibfnamefont {D.~E.}\ \bibnamefont
  {Soper}}, \bibinfo {author} {\bibfnamefont {M.}~\bibnamefont {Spannowsky}},
  \bibinfo {author} {\bibfnamefont {C.~J.}\ \bibnamefont {Wallace}}, \ and\
  \bibinfo {author} {\bibfnamefont {T.~M.~P.}\ \bibnamefont {Tait}},\ }\href
  {\doibase 10.1103/PhysRevD.90.115005} {\bibfield  {journal} {\bibinfo
  {journal} {Phys. Rev. D}\ }\textbf {\bibinfo {volume} {90}},\ \bibinfo
  {pages} {115005} (\bibinfo {year} {2014})},\ \Eprint
  {http://arxiv.org/abs/1407.2623} {arXiv:1407.2623 [hep-ph]} \BibitemShut
  {NoStop}%
\bibitem [{\citenamefont {Abe}\ \emph {et~al.}(1992)\citenamefont {Abe} \emph
  {et~al.}}]{Abe1992}%
  \BibitemOpen
  \bibfield  {author} {\bibinfo {author} {\bibfnamefont {F.}~\bibnamefont
  {Abe}} \emph {et~al.} (\bibinfo {collaboration} {CDF Collaboration}),\ }\href
  {\doibase 10.1103/PhysRevD.46.R1889} {\bibfield  {journal} {\bibinfo
  {journal} {Phys. Rev. D}\ }\textbf {\bibinfo {volume} {46}},\ \bibinfo
  {pages} {R1889} (\bibinfo {year} {1992})}\BibitemShut {NoStop}%
\bibitem [{\citenamefont {Buskulic}\ \emph {et~al.}(1993)\citenamefont
  {Buskulic} \emph {et~al.}}]{Buskulic1993}%
  \BibitemOpen
  \bibfield  {author} {\bibinfo {author} {\bibfnamefont {D.}~\bibnamefont
  {Buskulic}} \emph {et~al.},\ }\href@noop {} {\bibfield  {journal} {\bibinfo
  {journal} {Phys. Lett. B}\ }\textbf {\bibinfo {volume} {303}},\ \bibinfo
  {pages} {198} (\bibinfo {year} {1993})}\BibitemShut {NoStop}%
\bibitem [{\citenamefont {Akers}\ \emph {et~al.}(1995)\citenamefont {Akers}
  \emph {et~al.}}]{Akers1995}%
  \BibitemOpen
  \bibfield  {author} {\bibinfo {author} {\bibfnamefont {R.}~\bibnamefont
  {Akers}} \emph {et~al.},\ }\href@noop {} {\bibfield  {journal} {\bibinfo
  {journal} {Z. Phys. C}\ }\textbf {\bibinfo {volume} {67}},\ \bibinfo {pages}
  {203} (\bibinfo {year} {1995})}\BibitemShut {NoStop}%
\bibitem [{\citenamefont {Abreu}\ \emph {et~al.}(1997)\citenamefont {Abreu}
  \emph {et~al.}}]{Abreu1997}%
  \BibitemOpen
  \bibfield  {author} {\bibinfo {author} {\bibfnamefont {P.}~\bibnamefont
  {Abreu}} \emph {et~al.},\ }\href@noop {} {\bibfield  {journal} {\bibinfo
  {journal} {Phys. Lett. B}\ }\textbf {\bibinfo {volume} {396}},\ \bibinfo
  {pages} {315} (\bibinfo {year} {1997})}\BibitemShut {NoStop}%
\bibitem [{\citenamefont {Acosta}\ \emph {et~al.}(2003)\citenamefont {Acosta}
  \emph {et~al.}}]{Acosta2003}%
  \BibitemOpen
  \bibfield  {author} {\bibinfo {author} {\bibfnamefont {D.}~\bibnamefont
  {Acosta}} \emph {et~al.},\ }\href@noop {} {\bibfield  {journal} {\bibinfo
  {journal} {Phys. Rev. Lett.}\ }\textbf {\bibinfo {volume} {90}},\ \bibinfo
  {pages} {131801} (\bibinfo {year} {2003})}\BibitemShut {NoStop}%
\bibitem [{\citenamefont {Abbiendi}\ \emph {et~al.}(2003)\citenamefont
  {Abbiendi} \emph {et~al.}}]{Abbiendi2003}%
  \BibitemOpen
  \bibfield  {author} {\bibinfo {author} {\bibfnamefont {G.}~\bibnamefont
  {Abbiendi}} \emph {et~al.},\ }\href@noop {} {\bibfield  {journal} {\bibinfo
  {journal} {Phys. Lett. B}\ }\textbf {\bibinfo {volume} {572}},\ \bibinfo
  {pages} {8} (\bibinfo {year} {2003})}\BibitemShut {NoStop}%
\bibitem [{\citenamefont {Jaeckel}\ \emph {et~al.}(2013)\citenamefont
  {Jaeckel}, \citenamefont {Jankowiak},\ and\ \citenamefont
  {Spannowsky}}]{Jaeckel2012}%
  \BibitemOpen
  \bibfield  {author} {\bibinfo {author} {\bibfnamefont {J.}~\bibnamefont
  {Jaeckel}}, \bibinfo {author} {\bibfnamefont {M.}~\bibnamefont {Jankowiak}},
  \ and\ \bibinfo {author} {\bibfnamefont {M.}~\bibnamefont {Spannowsky}},\
  }\href {\doibase 10.1016/j.dark.2013.06.001} {\bibfield  {journal} {\bibinfo
  {journal} {Phys. Dark Univ.}\ }\textbf {\bibinfo {volume} {2}},\ \bibinfo
  {pages} {111} (\bibinfo {year} {2013})},\ \Eprint
  {http://arxiv.org/abs/1212.3620} {arXiv:1212.3620 [hep-ph]} \BibitemShut
  {NoStop}%
\bibitem [{\citenamefont {Chatrchyan}\ \emph {et~al.}(2013)\citenamefont
  {Chatrchyan} \emph {et~al.}}]{Chatrchyan2013}%
  \BibitemOpen
  \bibfield  {author} {\bibinfo {author} {\bibfnamefont {S.}~\bibnamefont
  {Chatrchyan}} \emph {et~al.} (\bibinfo {collaboration} {CMS Collaboration}),\
  }\href {\doibase 10.1103/PhysRevD.87.092008} {\bibfield  {journal} {\bibinfo
  {journal} {Phys. Rev. D}\ }\textbf {\bibinfo {volume} {87}},\ \bibinfo
  {pages} {092008} (\bibinfo {year} {2013})}\BibitemShut {NoStop}%
\bibitem [{\citenamefont {Dobroliubov}\ and\ \citenamefont
  {Ignatiev}(1990)}]{Dobroliubov1990}%
  \BibitemOpen
  \bibfield  {author} {\bibinfo {author} {\bibfnamefont {M.~I.}\ \bibnamefont
  {Dobroliubov}}\ and\ \bibinfo {author} {\bibfnamefont {A.~Y.}\ \bibnamefont
  {Ignatiev}},\ }\href {\doibase 10.1103/PhysRevLett.65.679} {\bibfield
  {journal} {\bibinfo  {journal} {Phys. Rev. Lett.}\ }\textbf {\bibinfo
  {volume} {65}},\ \bibinfo {pages} {679} (\bibinfo {year} {1990})}\BibitemShut
  {NoStop}%
\bibitem [{\citenamefont {Davidson}\ \emph {et~al.}(2000)\citenamefont
  {Davidson}, \citenamefont {Hannestad},\ and\ \citenamefont
  {Raffelt}}]{Davidson2000}%
  \BibitemOpen
  \bibfield  {author} {\bibinfo {author} {\bibfnamefont {S.}~\bibnamefont
  {Davidson}}, \bibinfo {author} {\bibfnamefont {S.}~\bibnamefont {Hannestad}},
  \ and\ \bibinfo {author} {\bibfnamefont {G.}~\bibnamefont {Raffelt}},\ }\href
  {\doibase 10.1088/1126-6708/2000/05/003} {\bibfield  {journal} {\bibinfo
  {journal} {JHEP}\ }\textbf {\bibinfo {volume} {05}},\ \bibinfo {pages} {003}
  (\bibinfo {year} {2000})},\ \Eprint {http://arxiv.org/abs/hep-ph/0001179}
  {arXiv:hep-ph/0001179 [hep-ph]} \BibitemShut {NoStop}%
\bibitem [{\citenamefont {Feng}\ \emph {et~al.}(2016)\citenamefont {Feng},
  \citenamefont {Smolinsky},\ and\ \citenamefont {Tanedo}}]{Feng2016}%
  \BibitemOpen
  \bibfield  {author} {\bibinfo {author} {\bibfnamefont {J.~L.}\ \bibnamefont
  {Feng}}, \bibinfo {author} {\bibfnamefont {J.}~\bibnamefont {Smolinsky}}, \
  and\ \bibinfo {author} {\bibfnamefont {P.}~\bibnamefont {Tanedo}},\ }\href
  {\doibase 10.1103/PhysRevD.93.115036} {\bibfield  {journal} {\bibinfo
  {journal} {Phys. Rev. D}\ }\textbf {\bibinfo {volume} {93}},\ \bibinfo
  {pages} {115036} (\bibinfo {year} {2016})}\BibitemShut {NoStop}%
\bibitem [{\citenamefont {Ahlers}(2009)}]{Ahlers2009}%
  \BibitemOpen
  \bibfield  {author} {\bibinfo {author} {\bibfnamefont {M.}~\bibnamefont
  {Ahlers}},\ }\href {\doibase 10.1103/PhysRevD.80.023513} {\bibfield
  {journal} {\bibinfo  {journal} {Phys. Rev. D}\ }\textbf {\bibinfo {volume}
  {80}},\ \bibinfo {pages} {023513} (\bibinfo {year} {2009})}\BibitemShut
  {NoStop}%
\bibitem [{\citenamefont {Burrage}\ \emph {et~al.}(2009)\citenamefont {Burrage}
  \emph {et~al.}}]{Burrage2009}%
  \BibitemOpen
  \bibfield  {author} {\bibinfo {author} {\bibfnamefont {C.}~\bibnamefont
  {Burrage}} \emph {et~al.},\ }\href
  {http://stacks.iop.org/1475-7516/2009/i=11/a=002} {\bibfield  {journal}
  {\bibinfo  {journal} {Journal of Cosmology and Astroparticle Physics}\ ,\
  \bibinfo {pages} {002}} (\bibinfo {year} {2009})}\BibitemShut {NoStop}%
\bibitem [{\citenamefont {Dubovsky}\ \emph {et~al.}(2004)\citenamefont
  {Dubovsky}, \citenamefont {Gorbunov},\ and\ \citenamefont
  {Rubtsov}}]{Dubovsky2003}%
  \BibitemOpen
  \bibfield  {author} {\bibinfo {author} {\bibfnamefont {S.~L.}\ \bibnamefont
  {Dubovsky}}, \bibinfo {author} {\bibfnamefont {D.~S.}\ \bibnamefont
  {Gorbunov}}, \ and\ \bibinfo {author} {\bibfnamefont {G.~I.}\ \bibnamefont
  {Rubtsov}},\ }\href {\doibase 10.1134/1.1675909} {\bibfield  {journal}
  {\bibinfo  {journal} {JETP Lett.}\ }\textbf {\bibinfo {volume} {79}},\
  \bibinfo {pages} {1} (\bibinfo {year} {2004})},\ \bibinfo {note} {[Pisma Zh.
  Eksp. Teor. Fiz.79,3(2004)]},\ \Eprint {http://arxiv.org/abs/hep-ph/0311189}
  {arXiv:hep-ph/0311189 [hep-ph]} \BibitemShut {NoStop}%
\bibitem [{\citenamefont {Berezhiani}\ and\ \citenamefont
  {Lepidi}(2009)}]{Berezhiani2009}%
  \BibitemOpen
  \bibfield  {author} {\bibinfo {author} {\bibfnamefont {Z.}~\bibnamefont
  {Berezhiani}}\ and\ \bibinfo {author} {\bibfnamefont {A.}~\bibnamefont
  {Lepidi}},\ }\href {\doibase https://doi.org/10.1016/j.physletb.2009.10.023}
  {\bibfield  {journal} {\bibinfo  {journal} {Physics Letters B}\ }\textbf
  {\bibinfo {volume} {681}},\ \bibinfo {pages} {276 } (\bibinfo {year}
  {2009})}\BibitemShut {NoStop}%
\bibitem [{\citenamefont {Vogel}\ and\ \citenamefont
  {Redondo}(2014)}]{Vogel2013}%
  \BibitemOpen
  \bibfield  {author} {\bibinfo {author} {\bibfnamefont {H.}~\bibnamefont
  {Vogel}}\ and\ \bibinfo {author} {\bibfnamefont {J.}~\bibnamefont
  {Redondo}},\ }\href {\doibase 10.1088/1475-7516/2014/02/029} {\bibfield
  {journal} {\bibinfo  {journal} {JCAP}\ }\textbf {\bibinfo {volume} {1402}},\
  \bibinfo {pages} {029} (\bibinfo {year} {2014})},\ \Eprint
  {http://arxiv.org/abs/1311.2600} {arXiv:1311.2600 [hep-ph]} \BibitemShut
  {NoStop}%
\bibitem [{\citenamefont {Dolgov}\ and\ \citenamefont
  {Rudenko}(2017)}]{Dolgov2017}%
  \BibitemOpen
  \bibfield  {author} {\bibinfo {author} {\bibfnamefont {A.~D.}\ \bibnamefont
  {Dolgov}}\ and\ \bibinfo {author} {\bibfnamefont {A.~S.}\ \bibnamefont
  {Rudenko}},\ }\href {\doibase 10.1134/S1063776117030116} {\bibfield
  {journal} {\bibinfo  {journal} {Journal of Experimental and Theoretical
  Physics}\ }\textbf {\bibinfo {volume} {124}},\ \bibinfo {pages} {564}
  (\bibinfo {year} {2017})}\BibitemShut {NoStop}%
\bibitem [{\citenamefont {Mohapatra}\ and\ \citenamefont
  {Rothstein}(1990)}]{Mohapatra1990}%
  \BibitemOpen
  \bibfield  {author} {\bibinfo {author} {\bibfnamefont {R.~N.}\ \bibnamefont
  {Mohapatra}}\ and\ \bibinfo {author} {\bibfnamefont {I.~Z.}\ \bibnamefont
  {Rothstein}},\ }\href {\doibase 10.1016/0370-2693(90)91907-S} {\bibfield
  {journal} {\bibinfo  {journal} {Phys. Lett. B}\ }\textbf {\bibinfo {volume}
  {247}},\ \bibinfo {pages} {593} (\bibinfo {year} {1990})}\BibitemShut
  {NoStop}%
\bibitem [{\citenamefont {Huang}\ \emph {et~al.}(2015)\citenamefont {Huang},
  \citenamefont {Zheng}, \citenamefont {Wang},\ and\ \citenamefont
  {Li}}]{Huang2015}%
  \BibitemOpen
  \bibfield  {author} {\bibinfo {author} {\bibfnamefont {X.}~\bibnamefont
  {Huang}}, \bibinfo {author} {\bibfnamefont {X.-P.}\ \bibnamefont {Zheng}},
  \bibinfo {author} {\bibfnamefont {W.-H.}\ \bibnamefont {Wang}}, \ and\
  \bibinfo {author} {\bibfnamefont {S.-Z.}\ \bibnamefont {Li}},\ }\href
  {\doibase 10.1103/PhysRevD.91.123513} {\bibfield  {journal} {\bibinfo
  {journal} {Phys. Rev. D}\ }\textbf {\bibinfo {volume} {91}},\ \bibinfo
  {pages} {123513} (\bibinfo {year} {2015})}\BibitemShut {NoStop}%
\bibitem [{\citenamefont {Korwar}\ and\ \citenamefont
  {Thalapillil}(2017)}]{Korwar2017}%
  \BibitemOpen
  \bibfield  {author} {\bibinfo {author} {\bibfnamefont {M.}~\bibnamefont
  {Korwar}}\ and\ \bibinfo {author} {\bibfnamefont {A.~M.}\ \bibnamefont
  {Thalapillil}},\ }\href@noop {} {} (\bibinfo {year} {2017}),\ \Eprint
  {http://arxiv.org/abs/arXiv:1709.07888} {arXiv:1709.07888} \BibitemShut
  {NoStop}%
\bibitem [{\citenamefont {Kvam}\ and\ \citenamefont
  {Latimer}(2014)}]{Kvam2014}%
  \BibitemOpen
  \bibfield  {author} {\bibinfo {author} {\bibfnamefont {A.~K.}\ \bibnamefont
  {Kvam}}\ and\ \bibinfo {author} {\bibfnamefont {D.~C.}\ \bibnamefont
  {Latimer}},\ }\href@noop {} {} (\bibinfo {year} {2014}),\ \Eprint
  {http://arxiv.org/abs/arXiv:1412.0708} {arXiv:1412.0708} \BibitemShut
  {NoStop}%
\bibitem [{\citenamefont {Kadota}\ \emph {et~al.}(2016)\citenamefont {Kadota},
  \citenamefont {Sekiguchia},\ and\ \citenamefont {Tashirob}}]{Kadota2016}%
  \BibitemOpen
  \bibfield  {author} {\bibinfo {author} {\bibfnamefont {K.}~\bibnamefont
  {Kadota}}, \bibinfo {author} {\bibfnamefont {T.}~\bibnamefont {Sekiguchia}},
  \ and\ \bibinfo {author} {\bibfnamefont {H.}~\bibnamefont {Tashirob}},\
  }\href@noop {} {} (\bibinfo {year} {2016}),\ \Eprint
  {http://arxiv.org/abs/arXiv:1602.04009} {arXiv:1602.04009} \BibitemShut
  {NoStop}%
\bibitem [{\citenamefont {Davidson}\ \emph {et~al.}(1991)\citenamefont
  {Davidson}, \citenamefont {Campbell},\ and\ \citenamefont
  {Bailey}}]{Davidson1991}%
  \BibitemOpen
  \bibfield  {author} {\bibinfo {author} {\bibfnamefont {S.}~\bibnamefont
  {Davidson}}, \bibinfo {author} {\bibfnamefont {B.}~\bibnamefont {Campbell}},
  \ and\ \bibinfo {author} {\bibfnamefont {D.~C.}\ \bibnamefont {Bailey}},\
  }\href {\doibase 10.1103/PhysRevD.43.2314} {\bibfield  {journal} {\bibinfo
  {journal} {Phys. Rev. D}\ }\textbf {\bibinfo {volume} {43}},\ \bibinfo
  {pages} {2314} (\bibinfo {year} {1991})}\BibitemShut {NoStop}%
\bibitem [{\citenamefont {Gl\"uck}\ \emph {et~al.}(2007)\citenamefont
  {Gl\"uck}, \citenamefont {Rakshit},\ and\ \citenamefont {Reya}}]{Gluck2007}%
  \BibitemOpen
  \bibfield  {author} {\bibinfo {author} {\bibfnamefont {M.}~\bibnamefont
  {Gl\"uck}}, \bibinfo {author} {\bibfnamefont {S.}~\bibnamefont {Rakshit}}, \
  and\ \bibinfo {author} {\bibfnamefont {E.}~\bibnamefont {Reya}},\ }\href
  {\doibase 10.1103/PhysRevD.76.091701} {\bibfield  {journal} {\bibinfo
  {journal} {Phys. Rev. D}\ }\textbf {\bibinfo {volume} {76}},\ \bibinfo
  {pages} {091701} (\bibinfo {year} {2007})}\BibitemShut {NoStop}%
\bibitem [{\citenamefont {Hu}\ \emph {et~al.}(2017)\citenamefont {Hu},
  \citenamefont {Kusenko},\ and\ \citenamefont {Takhistov}}]{Hu2017}%
  \BibitemOpen
  \bibfield  {author} {\bibinfo {author} {\bibfnamefont {P.-K.}\ \bibnamefont
  {Hu}}, \bibinfo {author} {\bibfnamefont {A.}~\bibnamefont {Kusenko}}, \ and\
  \bibinfo {author} {\bibfnamefont {V.}~\bibnamefont {Takhistov}},\ }\href
  {\doibase https://doi.org/10.1016/j.physletb.2017.02.035} {\bibfield
  {journal} {\bibinfo  {journal} {Physics Letters B}\ }\textbf {\bibinfo
  {volume} {768}},\ \bibinfo {pages} {18 } (\bibinfo {year}
  {2017})}\BibitemShut {NoStop}%
\bibitem [{\citenamefont {Badertscher}\ \emph {et~al.}(2007)\citenamefont
  {Badertscher}, \citenamefont {Crivelli}, \citenamefont {Fetscher},
  \citenamefont {Gendotti}, \citenamefont {Gninenko}, \citenamefont {Postoev},
  \citenamefont {Rubbia}, \citenamefont {Samoylenko},\ and\ \citenamefont
  {Sillou}}]{Badertscher2006}%
  \BibitemOpen
  \bibfield  {author} {\bibinfo {author} {\bibfnamefont {A.}~\bibnamefont
  {Badertscher}}, \bibinfo {author} {\bibfnamefont {P.}~\bibnamefont
  {Crivelli}}, \bibinfo {author} {\bibfnamefont {W.}~\bibnamefont {Fetscher}},
  \bibinfo {author} {\bibfnamefont {U.}~\bibnamefont {Gendotti}}, \bibinfo
  {author} {\bibfnamefont {S.~N.}\ \bibnamefont {Gninenko}}, \bibinfo {author}
  {\bibfnamefont {V.}~\bibnamefont {Postoev}}, \bibinfo {author} {\bibfnamefont
  {A.}~\bibnamefont {Rubbia}}, \bibinfo {author} {\bibfnamefont
  {V.}~\bibnamefont {Samoylenko}}, \ and\ \bibinfo {author} {\bibfnamefont
  {D.}~\bibnamefont {Sillou}},\ }\href {\doibase 10.1103/PhysRevD.75.032004}
  {\bibfield  {journal} {\bibinfo  {journal} {Phys. Rev. D}\ }\textbf {\bibinfo
  {volume} {75}},\ \bibinfo {pages} {032004} (\bibinfo {year} {2007})},\
  \Eprint {http://arxiv.org/abs/hep-ex/0609059} {arXiv:hep-ex/0609059 [hep-ex]}
  \BibitemShut {NoStop}%
\bibitem [{\citenamefont {Gninenko}\ \emph {et~al.}(2007)\citenamefont
  {Gninenko}, \citenamefont {Krasnikov},\ and\ \citenamefont
  {Rubbia}}]{Gninenko2006}%
  \BibitemOpen
  \bibfield  {author} {\bibinfo {author} {\bibfnamefont {S.~N.}\ \bibnamefont
  {Gninenko}}, \bibinfo {author} {\bibfnamefont {N.~V.}\ \bibnamefont
  {Krasnikov}}, \ and\ \bibinfo {author} {\bibfnamefont {A.}~\bibnamefont
  {Rubbia}},\ }\href {\doibase 10.1103/PhysRevD.75.075014} {\bibfield
  {journal} {\bibinfo  {journal} {Phys. Rev. D}\ }\textbf {\bibinfo {volume}
  {75}},\ \bibinfo {pages} {075014} (\bibinfo {year} {2007})},\ \Eprint
  {http://arxiv.org/abs/hep-ph/0612203} {arXiv:hep-ph/0612203 [hep-ph]}
  \BibitemShut {NoStop}%
\bibitem [{\citenamefont {Chen}\ \emph {et~al.}(2014)\citenamefont {Chen},
  \citenamefont {Chi}, \citenamefont {Li}, \citenamefont {Liu}, \citenamefont
  {Singh}, \citenamefont {Wong}, \citenamefont {Wu},\ and\ \citenamefont
  {Wu}}]{Chen2014}%
  \BibitemOpen
  \bibfield  {author} {\bibinfo {author} {\bibfnamefont {J.-W.}\ \bibnamefont
  {Chen}}, \bibinfo {author} {\bibfnamefont {H.-C.}\ \bibnamefont {Chi}},
  \bibinfo {author} {\bibfnamefont {H.-B.}\ \bibnamefont {Li}}, \bibinfo
  {author} {\bibfnamefont {C.-P.}\ \bibnamefont {Liu}}, \bibinfo {author}
  {\bibfnamefont {L.}~\bibnamefont {Singh}}, \bibinfo {author} {\bibfnamefont
  {H.~T.}\ \bibnamefont {Wong}}, \bibinfo {author} {\bibfnamefont {C.-L.}\
  \bibnamefont {Wu}}, \ and\ \bibinfo {author} {\bibfnamefont {C.-P.}\
  \bibnamefont {Wu}},\ }\href {\doibase 10.1103/PhysRevD.90.011301} {\bibfield
  {journal} {\bibinfo  {journal} {Phys. Rev. D}\ }\textbf {\bibinfo {volume}
  {90}},\ \bibinfo {pages} {011301} (\bibinfo {year} {2014})}\BibitemShut
  {NoStop}%
\bibitem [{\citenamefont {LaRue}\ \emph {et~al.}(1981)\citenamefont {LaRue},
  \citenamefont {Phillips},\ and\ \citenamefont {Fairbank}}]{LaRue1981}%
  \BibitemOpen
  \bibfield  {author} {\bibinfo {author} {\bibfnamefont {G.~S.}\ \bibnamefont
  {LaRue}}, \bibinfo {author} {\bibfnamefont {J.~D.}\ \bibnamefont {Phillips}},
  \ and\ \bibinfo {author} {\bibfnamefont {W.~M.}\ \bibnamefont {Fairbank}},\
  }\href {\doibase 10.1103/PhysRevLett.46.967} {\bibfield  {journal} {\bibinfo
  {journal} {Phys. Rev. Lett.}\ }\textbf {\bibinfo {volume} {46}},\ \bibinfo
  {pages} {967} (\bibinfo {year} {1981})}\BibitemShut {NoStop}%
\bibitem [{\citenamefont {Smith}\ \emph {et~al.}(1985)\citenamefont {Smith}
  \emph {et~al.}}]{Smith1985}%
  \BibitemOpen
  \bibfield  {author} {\bibinfo {author} {\bibfnamefont {P.}~\bibnamefont
  {Smith}} \emph {et~al.},\ }\href {\doibase
  http://dx.doi.org/10.1016/0370-2693(85)91426-1} {\bibfield  {journal}
  {\bibinfo  {journal} {Physics Letters B}\ }\textbf {\bibinfo {volume}
  {153}},\ \bibinfo {pages} {188 } (\bibinfo {year} {1985})}\BibitemShut
  {NoStop}%
\bibitem [{\citenamefont {Moore}\ \emph {et~al.}(2014)\citenamefont {Moore},
  \citenamefont {Rider},\ and\ \citenamefont {Gratta}}]{Moore2014}%
  \BibitemOpen
  \bibfield  {author} {\bibinfo {author} {\bibfnamefont {D.~C.}\ \bibnamefont
  {Moore}}, \bibinfo {author} {\bibfnamefont {A.~D.}\ \bibnamefont {Rider}}, \
  and\ \bibinfo {author} {\bibfnamefont {G.}~\bibnamefont {Gratta}},\ }\href
  {\doibase 10.1103/PhysRevLett.113.251801} {\bibfield  {journal} {\bibinfo
  {journal} {Phys. Rev. Lett.}\ }\textbf {\bibinfo {volume} {113}},\ \bibinfo
  {pages} {251801} (\bibinfo {year} {2014})}\BibitemShut {NoStop}%
\bibitem [{\citenamefont {Kim}\ \emph {et~al.}(2007)\citenamefont {Kim},
  \citenamefont {Lee}, \citenamefont {Lee}, \citenamefont {Perl}, \citenamefont
  {Halyo},\ and\ \citenamefont {Loomba}}]{Kim2007}%
  \BibitemOpen
  \bibfield  {author} {\bibinfo {author} {\bibfnamefont {P.~C.}\ \bibnamefont
  {Kim}}, \bibinfo {author} {\bibfnamefont {E.~R.}\ \bibnamefont {Lee}},
  \bibinfo {author} {\bibfnamefont {I.~T.}\ \bibnamefont {Lee}}, \bibinfo
  {author} {\bibfnamefont {M.~L.}\ \bibnamefont {Perl}}, \bibinfo {author}
  {\bibfnamefont {V.}~\bibnamefont {Halyo}}, \ and\ \bibinfo {author}
  {\bibfnamefont {D.}~\bibnamefont {Loomba}},\ }\href {\doibase
  10.1103/PhysRevLett.99.161804} {\bibfield  {journal} {\bibinfo  {journal}
  {Phys. Rev. Lett.}\ }\textbf {\bibinfo {volume} {99}},\ \bibinfo {pages}
  {161804} (\bibinfo {year} {2007})}\BibitemShut {NoStop}%
\bibitem [{\citenamefont {Halyo}\ \emph {et~al.}(2000)\citenamefont {Halyo},
  \citenamefont {Kim}, \citenamefont {Lee}, \citenamefont {Lee}, \citenamefont
  {Loomba},\ and\ \citenamefont {Perl}}]{Halyo2000}%
  \BibitemOpen
  \bibfield  {author} {\bibinfo {author} {\bibfnamefont {V.}~\bibnamefont
  {Halyo}}, \bibinfo {author} {\bibfnamefont {P.}~\bibnamefont {Kim}}, \bibinfo
  {author} {\bibfnamefont {E.~R.}\ \bibnamefont {Lee}}, \bibinfo {author}
  {\bibfnamefont {I.~T.}\ \bibnamefont {Lee}}, \bibinfo {author} {\bibfnamefont
  {D.}~\bibnamefont {Loomba}}, \ and\ \bibinfo {author} {\bibfnamefont {M.~L.}\
  \bibnamefont {Perl}},\ }\href {\doibase 10.1103/PhysRevLett.84.2576}
  {\bibfield  {journal} {\bibinfo  {journal} {Phys. Rev. Lett.}\ }\textbf
  {\bibinfo {volume} {84}},\ \bibinfo {pages} {2576} (\bibinfo {year}
  {2000})}\BibitemShut {NoStop}%
\bibitem [{\citenamefont {Ambrosio}\ \emph {et~al.}(2000)\citenamefont
  {Ambrosio} \emph {et~al.}}]{Ambrosio2000}%
  \BibitemOpen
  \bibfield  {author} {\bibinfo {author} {\bibfnamefont {M.}~\bibnamefont
  {Ambrosio}} \emph {et~al.},\ }\href@noop {} {\bibfield  {journal} {\bibinfo
  {journal} {Phys. Rev. D}\ }\textbf {\bibinfo {volume} {62}},\ \bibinfo
  {pages} {052003} (\bibinfo {year} {2000})}\BibitemShut {NoStop}%
\bibitem [{\citenamefont {Ambrosio}\ \emph {et~al.}(2004)\citenamefont
  {Ambrosio} \emph {et~al.}}]{Ambrosio2004}%
  \BibitemOpen
  \bibfield  {author} {\bibinfo {author} {\bibfnamefont {M.}~\bibnamefont
  {Ambrosio}} \emph {et~al.},\ }\href@noop {} {} (\bibinfo {year} {2004}),\
  \Eprint {http://arxiv.org/abs/arXiv:hep-ex/0402006} {arXiv:hep-ex/0402006}
  \BibitemShut {NoStop}%
\bibitem [{\citenamefont {Mori}\ \emph {et~al.}(1991)\citenamefont {Mori} \emph
  {et~al.}}]{Mori1991}%
  \BibitemOpen
  \bibfield  {author} {\bibinfo {author} {\bibfnamefont {M.}~\bibnamefont
  {Mori}} \emph {et~al.},\ }\href {\doibase 10.1103/PhysRevD.43.2843}
  {\bibfield  {journal} {\bibinfo  {journal} {Phys. Rev. D}\ }\textbf {\bibinfo
  {volume} {43}},\ \bibinfo {pages} {2843} (\bibinfo {year}
  {1991})}\BibitemShut {NoStop}%
\bibitem [{\citenamefont {Aglietta}\ \emph {et~al.}(1994)\citenamefont
  {Aglietta} \emph {et~al.}}]{Aglietta1994}%
  \BibitemOpen
  \bibfield  {author} {\bibinfo {author} {\bibfnamefont {M.}~\bibnamefont
  {Aglietta}} \emph {et~al.},\ }\href@noop {} {\bibfield  {journal} {\bibinfo
  {journal} {Astropart. Phys.}\ }\textbf {\bibinfo {volume} {2}},\ \bibinfo
  {pages} {29} (\bibinfo {year} {1994})}\BibitemShut {NoStop}%
\bibitem [{\citenamefont {Ahmed}\ \emph {et~al.}(2010)\citenamefont {Ahmed}
  \emph {et~al.}}]{CDMS2010}%
  \BibitemOpen
  \bibfield  {author} {\bibinfo {author} {\bibfnamefont {Z.}~\bibnamefont
  {Ahmed}} \emph {et~al.} (\bibinfo {collaboration} {CDMS II Collaboration}),\
  }\href {\doibase 10.1126/science.1186112} {\bibfield  {journal} {\bibinfo
  {journal} {Science}\ }\textbf {\bibinfo {volume} {327}},\ \bibinfo {pages}
  {1619} (\bibinfo {year} {2010})}\BibitemShut {NoStop}%
\bibitem [{\citenamefont {Agnese}\ \emph {et~al.}(2015)\citenamefont {Agnese}
  \emph {et~al.}}]{Agnese2015}%
  \BibitemOpen
  \bibfield  {author} {\bibinfo {author} {\bibfnamefont {R.}~\bibnamefont
  {Agnese}} \emph {et~al.} (\bibinfo {collaboration} {CDMS Collaboration}),\
  }\href {\doibase 10.1103/PhysRevLett.114.111302} {\bibfield  {journal}
  {\bibinfo  {journal} {Phys. Rev. Lett.}\ }\textbf {\bibinfo {volume} {114}},\
  \bibinfo {pages} {111302} (\bibinfo {year} {2015})}\BibitemShut {NoStop}%
\bibitem [{\citenamefont {Perl}\ \emph {et~al.}(2009)\citenamefont {Perl},
  \citenamefont {Lee},\ and\ \citenamefont {Loomba}}]{Perl2009}%
  \BibitemOpen
  \bibfield  {author} {\bibinfo {author} {\bibfnamefont {M.~L.}\ \bibnamefont
  {Perl}}, \bibinfo {author} {\bibfnamefont {E.~R.}\ \bibnamefont {Lee}}, \
  and\ \bibinfo {author} {\bibfnamefont {D.}~\bibnamefont {Loomba}},\
  }\href@noop {} {\bibfield  {journal} {\bibinfo  {journal} {Annu. Rev. Nucl.
  Part. Sci.}\ }\textbf {\bibinfo {volume} {59}},\ \bibinfo {pages} {47}
  (\bibinfo {year} {2009})}\BibitemShut {NoStop}%
\bibitem [{\citenamefont {Haas}\ \emph {et~al.}(2015)\citenamefont {Haas} \emph
  {et~al.}}]{Haas2015}%
  \BibitemOpen
  \bibfield  {author} {\bibinfo {author} {\bibfnamefont {A.}~\bibnamefont
  {Haas}} \emph {et~al.},\ }\href {arXiv:1410.6816} {\bibfield  {journal}
  {\bibinfo  {journal} {Phys. Lett. B}\ }\textbf {\bibinfo {volume} {746}},\
  \bibinfo {pages} {117} (\bibinfo {year} {2015})}\BibitemShut {NoStop}%
\bibitem [{\citenamefont {Abgrall}\ \emph {et~al.}(2014)\citenamefont {Abgrall}
  \emph {et~al.}}]{Abgrall2014}%
  \BibitemOpen
  \bibfield  {author} {\bibinfo {author} {\bibfnamefont {N.}~\bibnamefont
  {Abgrall}} \emph {et~al.},\ }\href {http://dx.doi.org/10.1155/2014/365432}
  {\bibfield  {journal} {\bibinfo  {journal} {Advances in High Energy Physics}\
  }\textbf {\bibinfo {volume} {2014}},\ \bibinfo {pages} {1} (\bibinfo {year}
  {2014})}\BibitemShut {NoStop}%
\bibitem [{\citenamefont {Aalseth}\ \emph {et~al.}(2017)\citenamefont {Aalseth}
  \emph {et~al.}}]{Abgrall2017c}%
  \BibitemOpen
  \bibfield  {author} {\bibinfo {author} {\bibfnamefont {C.}~\bibnamefont
  {Aalseth}} \emph {et~al.},\ }\href@noop {} {} (\bibinfo {year} {2017}),\
  \Eprint {http://arxiv.org/abs/https://arxiv.org/abs/1710.11608, accepted by
  PRL} {https://arxiv.org/abs/1710.11608, accepted by PRL} \BibitemShut
  {NoStop}%
\bibitem [{\citenamefont {Heise}(2015)}]{Heise2015}%
  \BibitemOpen
  \bibfield  {author} {\bibinfo {author} {\bibfnamefont {J.}~\bibnamefont
  {Heise}},\ }\href {\doibase 10.1088/1742-6596/606/1/012015} {\bibfield
  {journal} {\bibinfo  {journal} {Journal of Physics: Conference Series}\
  }\textbf {\bibinfo {volume} {606}},\ \bibinfo {pages} {012015} (\bibinfo
  {year} {2015})}\BibitemShut {NoStop}%
\bibitem [{\citenamefont {Abgrall}\ \emph
  {et~al.}(2017{\natexlab{a}})\citenamefont {Abgrall} \emph
  {et~al.}}]{Abgrall2017}%
  \BibitemOpen
  \bibfield  {author} {\bibinfo {author} {\bibfnamefont {N.}~\bibnamefont
  {Abgrall}} \emph {et~al.} (\bibinfo {collaboration} {{\sc Majorana}
  Collaboration}),\ }\href@noop {} {\bibfield  {journal} {\bibinfo  {journal}
  {Phys. Rev. Lett.}\ }\textbf {\bibinfo {volume} {118}},\ \bibinfo {pages}
  {161801} (\bibinfo {year} {2017}{\natexlab{a}})}\BibitemShut {NoStop}%
\bibitem [{\citenamefont {Luke}\ \emph {et~al.}(1989)\citenamefont {Luke},
  \citenamefont {Goulding}, \citenamefont {Madden},\ and\ \citenamefont
  {Pehl}}]{luk89}%
  \BibitemOpen
  \bibfield  {author} {\bibinfo {author} {\bibfnamefont {P.~N.}\ \bibnamefont
  {Luke}}, \bibinfo {author} {\bibfnamefont {F.~S.}\ \bibnamefont {Goulding}},
  \bibinfo {author} {\bibfnamefont {N.~W.}\ \bibnamefont {Madden}}, \ and\
  \bibinfo {author} {\bibfnamefont {R.~H.}\ \bibnamefont {Pehl}},\ }\href@noop
  {} {\bibfield  {journal} {\bibinfo  {journal} {IEEE Trans. Nucl. Sci.}\
  }\textbf {\bibinfo {volume} {36}},\ \bibinfo {pages} {926} (\bibinfo {year}
  {1989})}\BibitemShut {NoStop}%
\bibitem [{\citenamefont {Barbeau}\ \emph {et~al.}(2007)\citenamefont
  {Barbeau}, \citenamefont {Collar},\ and\ \citenamefont
  {Tench}}]{Barbeau2007}%
  \BibitemOpen
  \bibfield  {author} {\bibinfo {author} {\bibfnamefont {P.~S.}\ \bibnamefont
  {Barbeau}}, \bibinfo {author} {\bibfnamefont {J.~I.}\ \bibnamefont {Collar}},
  \ and\ \bibinfo {author} {\bibfnamefont {O.}~\bibnamefont {Tench}},\
  }\href@noop {} {\bibfield  {journal} {\bibinfo  {journal} {JCAP}\ }\textbf
  {\bibinfo {volume} {09}},\ \bibinfo {pages} {009} (\bibinfo {year}
  {2007})}\BibitemShut {NoStop}%
\bibitem [{\citenamefont {Aguayo}\ \emph {et~al.}(2011)\citenamefont {Aguayo}
  \emph {et~al.}}]{Aguayo2011}%
  \BibitemOpen
  \bibfield  {author} {\bibinfo {author} {\bibfnamefont {E.}~\bibnamefont
  {Aguayo}} \emph {et~al.},\ }\href@noop {} {} (\bibinfo {year} {2011}),\
  \bibinfo {note} {sLAC eConf C110809},\ \Eprint
  {http://arxiv.org/abs/arXiv:1109.6913} {arXiv:1109.6913} \BibitemShut
  {NoStop}%
\bibitem [{\citenamefont {Cooper}\ \emph {et~al.}(2011)\citenamefont {Cooper}
  \emph {et~al.}}]{Cooper2011}%
  \BibitemOpen
  \bibfield  {author} {\bibinfo {author} {\bibfnamefont {R.}~\bibnamefont
  {Cooper}} \emph {et~al.},\ }\href
  {http://www.sciencedirect.com/science/article/pii/S0168900210024915}
  {\bibfield  {journal} {\bibinfo  {journal} {NIM A}\ }\textbf {\bibinfo
  {volume} {629}},\ \bibinfo {pages} {303 } (\bibinfo {year}
  {2011})}\BibitemShut {NoStop}%
\bibitem [{\citenamefont {Feldman}\ and\ \citenamefont
  {Cousins}(1998)}]{Feldman1998}%
  \BibitemOpen
  \bibfield  {author} {\bibinfo {author} {\bibfnamefont {G.~J.}\ \bibnamefont
  {Feldman}}\ and\ \bibinfo {author} {\bibfnamefont {R.~D.}\ \bibnamefont
  {Cousins}},\ }\href {\doibase 10.1103/PhysRevD.57.3873} {\bibfield  {journal}
  {\bibinfo  {journal} {Phys. Rev. D}\ }\textbf {\bibinfo {volume} {57}},\
  \bibinfo {pages} {3873} (\bibinfo {year} {1998})}\BibitemShut {NoStop}%
\bibitem [{\citenamefont {Bauer}\ \emph {et~al.}(2006)\citenamefont {Bauer}
  \emph {et~al.}}]{Bauer2006}%
  \BibitemOpen
  \bibfield  {author} {\bibinfo {author} {\bibfnamefont {M.}~\bibnamefont
  {Bauer}} \emph {et~al.},\ }\href
  {http://stacks.iop.org/1742-6596/39/i=1/a=097} {\bibfield  {journal}
  {\bibinfo  {journal} {Journal of Physics: Conference Series}\ }\textbf
  {\bibinfo {volume} {39}},\ \bibinfo {pages} {362} (\bibinfo {year}
  {2006})}\BibitemShut {NoStop}%
\bibitem [{\citenamefont {Boswell}\ \emph {et~al.}(2011)\citenamefont {Boswell}
  \emph {et~al.}}]{Boswell2011}%
  \BibitemOpen
  \bibfield  {author} {\bibinfo {author} {\bibfnamefont {M.}~\bibnamefont
  {Boswell}} \emph {et~al.},\ }\href@noop {} {\bibfield  {journal} {\bibinfo
  {journal} {IEEE Trans. Nucl. Sci.}\ }\textbf {\bibinfo {volume} {58}},\
  \bibinfo {pages} {1212} (\bibinfo {year} {2011})}\BibitemShut {NoStop}%
\bibitem [{\citenamefont {Agostinelli}\ \emph {et~al.}(2003)\citenamefont
  {Agostinelli} \emph {et~al.}}]{Agostinelli2003}%
  \BibitemOpen
  \bibfield  {author} {\bibinfo {author} {\bibfnamefont {S.}~\bibnamefont
  {Agostinelli}} \emph {et~al.},\ }\href
  {http://www.sciencedirect.com/science/article/pii/S0168900203013688}
  {\bibfield  {journal} {\bibinfo  {journal} {NIM A}\ }\textbf {\bibinfo
  {volume} {506}},\ \bibinfo {pages} {250 } (\bibinfo {year}
  {2003})}\BibitemShut {NoStop}%
\bibitem [{\citenamefont {Prasad}(2013)}]{Prasad2013}%
  \BibitemOpen
  \bibfield  {author} {\bibinfo {author} {\bibfnamefont {K.~B.}\ \bibnamefont
  {Prasad}},\ }\href@noop {} {\bibfield  {journal} {\bibinfo  {journal} {PhD
  thesis, Texas A\&M University}\ } (\bibinfo {year} {2013})}\BibitemShut
  {NoStop}%
\bibitem [{\citenamefont {Allison}\ and\ \citenamefont
  {Cobb}(1980)}]{Allison1980}%
  \BibitemOpen
  \bibfield  {author} {\bibinfo {author} {\bibfnamefont {W.}~\bibnamefont
  {Allison}}\ and\ \bibinfo {author} {\bibfnamefont {J.}~\bibnamefont {Cobb}},\
  }\href@noop {} {\bibfield  {journal} {\bibinfo  {journal} {Ann. Rev. Nucl.
  Part. Sci.}\ }\textbf {\bibinfo {volume} {30}},\ \bibinfo {pages} {253}
  (\bibinfo {year} {1980})}\BibitemShut {NoStop}%
\bibitem [{\citenamefont {Bichsel}(2006)}]{Bichsel2006}%
  \BibitemOpen
  \bibfield  {author} {\bibinfo {author} {\bibfnamefont {H.}~\bibnamefont
  {Bichsel}},\ }\href@noop {} {\bibfield  {journal} {\bibinfo  {journal} {Nucl.
  Instrum. Meth. A}\ }\textbf {\bibinfo {volume} {562}},\ \bibinfo {pages}
  {154} (\bibinfo {year} {2006})}\BibitemShut {NoStop}%
\bibitem [{\citenamefont {Bichsel}(1988)}]{Bichsel1988}%
  \BibitemOpen
  \bibfield  {author} {\bibinfo {author} {\bibfnamefont {H.}~\bibnamefont
  {Bichsel}},\ }\href@noop {} {\bibfield  {journal} {\bibinfo  {journal} {Rev.
  Mod. Phys.}\ }\textbf {\bibinfo {volume} {60}},\ \bibinfo {pages} {663}
  (\bibinfo {year} {1988})}\BibitemShut {NoStop}%
\bibitem [{\citenamefont {Abgrall}\ \emph
  {et~al.}(2017{\natexlab{b}})\citenamefont {Abgrall} \emph
  {et~al.}}]{Abgrall2017b}%
  \BibitemOpen
  \bibfield  {author} {\bibinfo {author} {\bibfnamefont {N.}~\bibnamefont
  {Abgrall}} \emph {et~al.},\ }\href@noop {} {\bibfield  {journal} {\bibinfo
  {journal} {Astroparticle Physics}\ }\textbf {\bibinfo {volume} {93}},\
  \bibinfo {pages} {70} (\bibinfo {year} {2017}{\natexlab{b}})}\BibitemShut
  {NoStop}%
\end{thebibliography}%

\end{document}